\documentclass[aps,prb,twocolumn,amsmath,amssymb,superscriptaddress,floatfix,scrartcl]{revtex4-1}

\usepackage{amssymb}
\usepackage{color}
\usepackage{graphicx}
\usepackage{srcltx}
\usepackage{mathrsfs}
\usepackage[unicode=true,pdfusetitle,
 bookmarks=false,colorlinks=true,citecolor=blue,urlcolor=blue,linkcolor=red]{hyperref}

\begin{document}

\title{
Gauging (3+1)-dimensional topological phases:
an approach from surface theories
   }

\author{Xiao Chen}
\affiliation{
Kavli Institute for Theoretical Physics,
University of California at Santa Barbara, CA 93106, USA}

\author{Apoorv Tiwari}
\affiliation{
Institute for Condensed Matter Theory and
Department of Physics, University of Illinois
at Urbana-Champaign,
1110 West Green St, Urbana IL 61801, USA
            }

\author{Chetan Nayak}
\affiliation{
Station Q, Microsoft Research, Santa Barbara, California 93106-6105, USA
}
\affiliation{
Physics Department, University of California, Santa Barbara, CA 93106, USA
}

\author{Shinsei Ryu}
\affiliation{James Franck Institute and Kadanoff Center for Theoretical Physics,
University of Chicago, Illinois 60637, USA}

\date{\today}

\begin{abstract}
We discuss several bosonic topological phases in (3+1) dimensions
enriched by a global $\mathbb{Z}_2$ symmetry,
and gauging the $\mathbb{Z}_2$ symmetry.
More specifically,
following the spirit of the bulk-boundary correspondence,
expected to hold in topological phases of matter in general,
we consider boundary (surface) field theories
and their orbifold. 
From the surface partition functions, 
we extract the modular $\mathcal{S}$ and $\mathcal{T}$ matrices and compare them with $(2+1)$d toplogical phase after dimensional reduction.
As a specific example,
we discuss
topologically ordered phases in $(3+1)$ dimensions described by the BF topological quantum field theories,
with abelian exchange statistics between
point-like and loop-like quasiparticles.
Once the $\mathbb{Z}_2$ charge conjugation symmetry is gauged, the $\mathbb{Z}_2$ flux becomes non-abelian excitation.
The gauged topological phases we are considering here belong to the quantum double model with non-abelian group in $(3+1)$ dimensions.
\end{abstract}


\maketitle

\tableofcontents

\section{Introduction}

Symmetry and topology intertwine in many phases of matter.
Two prime examples are symmetry-protected topological (SPT) phases and
symmetry-enriched topological (SET) phases.
In SPT phases,
symmetries play a crucial role,
in that they are
sharply distinct from trivial phases (i.e., product states)
only in the presence of symmetries. \cite{HasanKane10,QiZhangreview11,Senthil2014,ChenGuLiuWen11,RMP}
On the other hand,
topologically ordered phases can be enriched by global symmetries.
The seminal example is the charge fractionalization
of Laughlin quasiparticles
in the fractional quantum Hall effect. The relevant global symmetry
in this case is $U(1)$ associated to
particle number conservation.
Varieties of SET phases
have been discussed in the literature.\cite{Wilczekbook, Fradkinbook, Wenbook}

A global symmetry in SPT or SET phases can be promoted
to a local symmetry through gauging.
Such ``gauging'' (in the bulk) or ``orbifolding'' (on the edge) is a useful tool
to understand parent SPT and SET phases in $(2+1)$ dimensions.\cite{Levin2012, Ryu2012, Sule2013,Hsieh2014} Gauging an SPT phase leads to a topological phase;
thus the SPT phase is the parent of this topological phase.
The topological class of the parent SPT phase can be inferred from,
and, in fact, has one-to-one correspondence with the topological order
(i.e., properties of anyons) which arises by the gauging procedure.
We can also gauge a global (discrete) symmetry $\mathcal{G}$ in an SET phase,
thereby giving rise to a new topological phase.
In particular, if the global symmetry acts on emergent excitations (anyons) by
permuting the anyon labels\cite{Khan2014} in the parent SET, then
gauging these symmetries will lead to more interesting "twist liquids".
\cite{Barkeshli_gauging_2014, Teo2015,Tarantino2016, Teotwistdefectreview}

The focus of this paper
is to generalize the above idea to $(3+1)$ dimensions and discuss
gauging/orbifolding global symmetries in $(3+1)$-dimensional bosonic
topologically ordered phases.
Starting from $(3+1)$-dimensional topologically ordered phases
with Abelian topological order,
which are described by (multi-component) BF theories,
we gauge a global $\mathbb{Z}_2$ symmetry 
and show that the new topological phase is non-Abelian and related to a non-Abelian quantum double model.

Previous work has constructed line defects in $(3+1)$-dimensional topological phases.\cite{Mesaros2013b,Bi2014,Ye2016} These semi-classical defects are analogous to twist defects in $(2+1)$d topological phases and can twist the anyon labels.\cite{Kitaev06, Bombin2010, KitaevKong12, YouWen, BarkeshliQi, BarkeshliJianQi, Teo2014, teo2013braiding} In this paper, we will fully gauge the discrete global symmetries so that these topological defects will become fully deconfined
loop-like excitations.
Our method for determining the resulting $(3+1)$d topological phase
relies on the bulk-boundary correspondence.
We work with the $(2+1)$-dimensional surface theories
of the bulk $(3+1)$d BF theories,
and consider the $\mathbb{Z}_2$ orbifold thereof.
As in the context of $(1+1)$-dimensional conformal
field theories (CFT), orbifolding a CFT amounts to
considering the partition functions in the presence of
twisted boundary conditions. \cite{Moore_Seiberg_1989, Dijkgraaf1989,GINSPARG1988,FMS-CFT}
By putting the surface theory on the spacetime torus $T^3$
with proper boundary conditions in time and two spatial directions,
we derive the transformation properties of
the twisted partition function under the mapping class group
(the large diffeomorphisms) of $T^3$.\cite{Hsieh2015, chen2015bulk} This procedure
allows us to read off the modular $\mathcal{S}$ and $\mathcal{T}$
matrices, which encode the properties of anyons
in the gauged surface theory. This, in turn, allows us to deduce the gauged bulk theory.


The rest of the paper is organized as follows:
In Sec.~\ref{orbifold_2d}, we briefly review orbifold CFTs in $(1+1)$d. We further elucidate this
with a simple example in the $\mathbb{Z}_{\mathrm{K}}$ quantum double model in Sec.~\ref{orbifold_z_k}. In Sec.~\ref{orbifold_3d}, we consider three different topological phases in $(3+1)$d and study the corresponding orbifolded surface theories. In Sec.~\ref{Discussion}, we discuss the bulk non-Abelian topological phase that results from
gauging $\mathbb{Z}_2$ symmetry and make a connection
with the surface orbifold theory. We summarize our results in Sec.~\ref{conclusion}. 

\section{Orbifolding the $(1+1)$d boundary theory}
\label{Orbifolding 1+1 boundary theory}

\subsection{Summary of orbifold CFT in $(1+1)$d}
\label{orbifold_2d}
Topologically ordered phases in $(2+1)$d
are often equipped (or enriched) with some global discrete symmetries.
These symmetries may permute anyon labels,
but leave the $\mathcal{S}$ and $\mathcal{T}$ matrices invariant.
Gauging the discrete symmetry $\mathcal{G}$
will give rise to more interesting topological phases.

A way to understand this new topological phase is through studying the boundary theory.
For a topological phase, the gapless boundary state can be described by a rational CFT $\mathcal{C}$.
Gauging the global symmetry in the bulk corresponds to orbifolding the edge CFT
by the symmetry.
By orbifolding by symmetry $\mathcal{G}$,
we project out the symmetry non-invariant states and, simultaneously,
add some twist sectors to the Hilbert space.
For the details of orbifold CFTs, see Ref.\ \onlinecite{Moore_Seiberg_1989, Dijkgraaf1989,GINSPARG1988,FMS-CFT}.
The orbifold CFT $\mathcal{C}/\mathcal{G}$ can be understood by
calculating the character for each primary field.
The characters are the partition functions under the symmetry projection,
\begin{align}
\chi^{h}_{n}=\mbox{Tr}_h( \mathcal{P}_{n}e^{-tH} )
\label{char_def}
\end{align}
where $\mathcal{P}_{n}$ is a projection operator, and
$h\in\mathcal{G}$ defines the twist in the spatial direction.
Here, 
the projection operator $\mathcal{P}_{n}$ depends on 
the set of phase factors $\{\omega\}$ known as
the discrete torsion phases.
E.g., 
if $\mathcal{G}$ is an abelian $\mathbb{Z}_N$ symmetry, the projection operator
is simply written as $\mathcal{P}_{n}=\sum_{k=0}^{N-1}\omega^{-nk}g^k/N$ where
$g$ is the generator of $\mathbb{Z}_N$
and
$\omega$ is the $k$-th root of unity.
The character $\chi^h_{n}$ 
can be understood as a linear combination of 
partition functions with fixed twist $h$ in the spatial direction
and all allowed twists $g$ in the time direction. 
Here, if $\mathcal{G}$ is a non-Abelian group, we require $g$ belongs to
centralizer subgroup of $h$ containing all $g$
that commutes with $h$, i.e., $gh=hg$.

The characters form a complete basis for a reprsentation of the group of modular transformations.
The modular $\mathcal{S}$ and $\mathcal{T}$ matrices for the orbifold CFT,
which are identical to the $\mathcal{S}$ and $\mathcal{T}$ matrices for the bulk topological order,
thus encode topological information of the topological phase.

One interesting example is the toric code model.
It has a global duality symmetry which exchanges the charge $e$ and flux $m$,
while leaving $\psi=e\times m$ invariant.
Gauging the $\mathbb{Z}_2$ duality symmetry leads to a non-Abelian topological phase.
The calculation of the characters for the edge orbifold CFT shows that $\mathcal{S}$ and $\mathcal{T}$ matrices are equivalent to that for the $\mbox{Ising}\times \overline{\mbox{Ising}}$ CFT and suggests that the bulk topological phase has nine quasi-particles, including the Ising-like anyon excitation.\cite{kitaev2006topological, Bais2009}

\subsection{Orbifolding $\mathbb{Z}_2$ symmetry on the boundary of the $\mathbb{Z}_{\mathrm{K}}$ quantum double model}
\label{orbifold_z_k}

Starting from an abelian topological phase,
if we gauge a global symmetry $G$,
the resulting non-abelian topological
phase includes the following three types of excitations:\cite{Teo2015}
First, for the anyon in {\bf a}
which is invariant under symmetry $G$,
they remain as excitations in the twist liquid.
They can also couple with gauge charge and form a composite
particle.
These excitations are abelian excitations with zero gauge flux and
called Type $(1)$ excitations.
Next, for anyons $\{ {\bf a}_i\}$ which are not invariant under symmetry,
they need to group together to form a superselection sector
so that ${\bf a}_1+{\bf a}_2+\ldots$ is gauge invariant.
Such Type (2) excitations are non-abelian quasi-particles with zero gauge flux.
Finally, Type $(3)$ excitations are the most interesting ones:
They carry non-trivial gauge flux and correspond
to non-abelian twist defects before gauging.
From the boundary field theory point of view, 
for these three types of excitations,
we can construct the corresponding characters on the boundary.

We now work out an example completely explicitly in order
to illuminate the general strategy:
$\mathbb{Z}_2$ charge-conjugation symmetry in
the $D(\mathbb{Z}_{\mathrm{K}})$ quantum double model.
We here consider the case when $\mathrm{K}$ is an odd number. The detail for this method can be found in Ref.\ \onlinecite{2d-prep}.

Our use of a gapless edge CFT deserves further comment.
For a single-component Chern-Simons theory, there is a chiral gapless mode on the boundary, which is stable and cannot be gapped out. This is because the single component Chern-Simons theory is anomalous and requires a gapless edge mode on the boundary to compensate the anomaly in the bulk.
Meanwhile, for the non-chiral $D(\mathbb{Z}_{\mathrm{K}})$ quantum double model without anomaly, the boundary CFT can be gapped out if we do not impose any symmetry.
Although the gapless CFT is not stable, it does encode topological data in the bulk.
Thus, we can use the ``fine-tuned'' gapless CFT as an intermediate step to study the bulk topological phase via the bulk boundary correspondence.
This is also true for the $(3+1)$d topological phase.

The $D(\mathbb{Z}_{\mathrm{K}})$ quantum double model has
two fundamental quasi-particle excitations,
$e$ and $m$.
All the quasi-particle excitations can be written as
$e^am^b$, where $0\leq a,b<\mathrm{K}$.
$e$ and $m$ are self-bosons,
and have non-trivial mutual braiding statistics
with braiding phase $e^{2\pi i/\mathrm{K}}$.

The $D(\mathbb{Z}_{\mathrm{K}})$ quantum double model has a global $\mathbb{Z}_2$ charge-conjugation symmetry which exchanges
$e^{a}m^b$ and $e^{\mathrm{K}-a}m^{\mathrm{K}-b}$.
If $\mathrm{K}$ is an odd number, there is no quasi-particle which is invariant
under the charge-conjugation.

Once the $\mathbb{Z}_2$ symmetry is gauged,
i.e., the global charge-conjugation symmetry is promoted
to a local gauge symmetry,
there is a $\mathbb{Z}_2$ bosonic charge $j$ which satisfies
the fusion rule $j\times j=1$.
On the other hand,
the $\mathbb{Z}_2$ flux $\sigma$ is a non-Abelian quasi-particle.
$\sigma$ can combine with $\mathbb{Z}_2$ charge to form the
flux-charge composite quasi-particle, $\tau=\sigma\times j$.
The original abelian anyons $e^am^b$ will group together to form gauge invariant superselection sector $e^am^b+e^{\mathrm{K}-a}m^{\mathrm{K}-b}$ with quantum dimension equal to 2.

Let us now take a look at the gauging procedure from the boundary field
theory point of view.
The relevant boundary theory is described by the
following Lagrangian density
\begin{align}
\mathcal{L}=\frac{1}{4\pi}(\partial_t\phi^I\mathbf{K}_{IJ}\partial_x\phi^J+
\partial_x \phi^IV_{IJ}\partial_x \phi^J),
\label{lutt}
\end{align}
where $(t,x)$ are the spacetime coordinates of the boundary;
$\mathbf{K}=\mathrm{K}\sigma_x$, $\vec{\Phi}=(\phi^1,\phi^2)$ is a two-component boson,
and $V$ is a symmetric and positive definite matrix that accounts
for the interaction on the edge and is non-universal.
This model has $\mathrm{K}^2$ characters and there is a choice of interaction $V_{IJ}$
for which they take the form
\begin{align}
B^{\mathrm{K}}_{a,b}(\tau)=
\frac{1}{|\eta(\tau)|^2}\sum_{s,t\in\mathbb{Z}}
q^{\frac{1}{4\mathrm{K}}(\mathrm{K}s+a+\mathrm{K}t+b)^2}
\bar{q}^{\frac{1}{4\mathrm{K}}(\mathrm{K}t+a-\mathrm{K}s-b)^2}
\label{toric_char}
\end{align}
where $a,b\in\mathbb{Z}\ \mbox{mod}\ \mathrm{K}$ are the anyon labels,
$\tau$ is the modular parameter of the spacetime torus,
$q=\exp (2\pi i\tau)$, and $\eta(\tau)$ is the Dedekind eta function.
The details of the calculation can be found in Appendix.

The boundary theory \eqref{lutt} is invariant under
the $\mathbb{Z}_2$ charge-conjugation symmetry
\begin{align}
\phi^{1,2} \to - \phi^{1,2}.
\end{align}
Once orbifolded by the $\mathbb{Z}_2$ symmetry,
$\phi^{1,2}$ can become $-\phi^{1,2}$ when the coordinates are taken around
the time and spatial directions on the $(1+1)$d torus.
Therefore orbifolding introduces anti-periodic boundary conditions
in the $x$ and $t$ directions.
The partition function with twisted boundary condition is labelled by $Z^{\mu\nu}$,
where $\mu, \nu =0,\frac{1}{2}$ represents untwisted and twisted boundary
condition in, respectively, the time and space directions.
The twisted partition functions are given by
\begin{align}
Z^{\frac{1}{2},0}=\left|\frac{\eta}{\theta_2}\right|,\quad Z^{0, \frac{1}{2}}=\left|\frac{\eta}{\theta_4}\right|,\quad Z^{\frac{1}{2},\frac{1}{2}}=\left|\frac{\eta}{\theta_3}\right|,
\end{align}
where $\theta_{2,3,4}$ are Jacobi theta functions
defined by
\begin{align}
  \theta_2=\sum_{n\in\mathbb{Z}}q^{\frac{1}{2}(n+\frac{1}{2})^2},\
  \theta_3=\sum_{n\in\mathbb{Z}}q^{\frac{n^2}{2}},\
  \theta_4=\sum_{n\in\mathbb{Z}}(-1)^nq^{\frac{n^2}{2}}.
\end{align}
One can readily check that under modular transformations,
\begin{align}
  & Z^{\frac{1}{2},0}\overset{\mathcal{S}}{\longleftrightarrow}Z^{0,\frac{1}{2}},
    \quad
    Z^{\frac{1}{2},\frac{1}{2}}\overset{\mathcal{S}}{\longleftrightarrow}Z^{\frac{1}{2},\frac{1}{2}},
    \nonumber\\
&
Z^{\frac{1}{2},0}\overset{\mathcal{T}}{\longleftrightarrow}Z^{\frac{1}{2},0},\quad Z^{0, \frac{1}{2}}\overset{\mathcal{T}}{\longleftrightarrow}Z^{\frac{1}{2},\frac{1}{2}}.
\end{align}

\begin{table}[t]
\centering
\begin{ruledtabular}
\begin{tabular}{lccc}
character $\chi$ & $d_\chi$ & $h_\chi$
& $\mathcal{N}$\\ \hline
$\chi_I=\frac{1}{2}B^{\mathrm{K}}_{0,0}+\frac{1}{2}\left|\frac{\eta}{\theta_2}\right|$ & $1$ & $0$ & $1$\\
$\chi_j=\frac{1}{2}B^{\mathrm{K}}_{0,0}-\frac{1}{2}\left|\frac{\eta}{\theta_2}\right|$ & $1$ & $0$ & $1$\\
$\chi_{a,b}=\frac{1}{2}B^{\mathrm{K}}_{a,b}+\frac{1}{2}B^{\mathrm{K}}_{\mathrm{K}-a,\mathrm{K}-b}$ & $2$ & $\frac{ab}{\mathrm{K}}$ & $\frac{\mathrm{K}^2-1}{2}$\\
$\chi_{\sigma}=\frac{1}{2}\left|\frac{\eta}{\theta_4}\right|+\frac{1}{2}\left|\frac{\eta}{\theta_3}\right|$ & $\mathrm{K}$ & $0$ & $1$\\
$\chi_{\tau}=\frac{1}{2}\left|\frac{\eta}{\theta_4}\right|-\frac{1}{2}\left|\frac{\eta}{\theta_3}\right|$ & $\mathrm{K}$ & $\frac{1}{2}$ & $1$\\
\end{tabular}
\end{ruledtabular}
\caption{
The quantum dimensions $d_\chi$,
conformal dimensions $h_{\chi}$,
and the number $\mathcal{N}$ of
characters $\chi$ from orbifolding
the charge-conjugate $\mathbb{Z}_2$ symmetry of Eq.\ \eqref{lutt} when $\mathrm{K}$ is odd.
The conformal dimensions $h_{\chi}$ are defined mod $\mathbb{Z}$.
For $\chi_{a,b}$, if $a\neq b$, we require $0\leq a<b\leq \mathrm{K}$ here.
If $a=b$, we require $a\leq (\mathrm{K}-1)/2$.}
\label{cc_z2_odd}
\end{table}

We use these twist blocks and the original characters
$B_{a,b}^{\mathrm{K}}$ to construct the characters for the orbifold CFT and the result is shown in Table \ref{cc_z2_odd}.
From the table we can see that the quantum dimension for a $\mathbb{Z}_2$ flux $\sigma$ is $\mathrm{K}$, indicating that it is a non-Abelian quasi-particle.
The $\mathcal{S}$ matrix is
\begin{align}
\mathcal{S}=
\frac{1}{\mathcal{D}}\begin{pmatrix}
1&1&2&\mathrm{K}&\mathrm{K}\\
1&1&2&-\mathrm{K}&-\mathrm{K}\\
2&2&4\cos[\frac{2\pi}{\mathrm{K}}(ab^{\prime}+ba^{\prime})]&0&0\\
\mathrm{K}&-\mathrm{K}&0&\mathrm{K}&-\mathrm{K}\\
\mathrm{K}&-\mathrm{K}&0&-\mathrm{K}&\mathrm{K}
\end{pmatrix}
\end{align}
where the total quantum dimension is $\mathcal{D}=2\mathrm{K}$. 
The topologica spin $e^{2\pi i h}$ of anyonic excitations can be read off 
from the (eigenvalues of the) $\mathcal{T}$ matrix.

A trivial example is $\mathrm{K}=1$.
Before gauging, the bulk has no topological order. 
After gauging the $\mathbb{Z}_2$ symmetry, the $\mathcal{S}$ and $\mathcal{T}$ matrix is the same as that for the toric code model, indicating the original phase is not a symmetry protected topological phase (SPT).\cite{Levin2012} 
The twist fields $\sigma,~\tau$ have 
quantum dimension equal to one, 
and correspond to the Abelian $\mathbb{Z}_2$ flux in the bulk.
For $\mathrm{K}\geq 3$, the gauged system has non-Abelian topological order
in the bulk. This non-abelian topological order can be described by
the $D(\mathcal{D}_{\mathrm{K}})$ quantum double model, 
where $\mathcal{D}_{\mathrm{K}}$ is the dihedral group of order $2\mathrm{K}$.
\cite{Tarantino2016}

\section{Gauging $\mathbb{Z}_2$ charge-conjugation symmetry
in (3+1)d $\mathbb{Z}_{\mathrm{K}}$ gauge theories}
\label{orbifold_3d}
In this section,
we discuss gauging discrete symmetries in $(3+1)$d topologically ordered phases.
A specific example we consider in this section
is the topological $\mathbb{Z}_{\mathrm{K}}$ gauge theory. 
It is 
the long wavelength limit of the deconfined phase of the $\mathbb{Z}_{\mathrm{K}}$ gauge theory.
A convenient description of the topological $\mathbb{Z}_{\mathrm{K}}$ 
gauge theory is given by 
the single-component $(3+1)$d BF theory,
which is defined by the following action 
\begin{align}
S_{bulk}
&=
\frac{\mathrm{K}}{2\pi}
\int_{\mathcal M}
b\wedge da
\label{BF bulk action}
\end{align}
where $\mathcal{M}$ is a (3+1)d spacetime manifold;
$a$ and $b$ are a one- and two-form, respectively;
$\mathrm{K}$ is an integral parameter (``level'').
This action describes the simplest topological phase in $(3+1)$d; its fundamental excitations are a
particle excitation $e$ and a loop excitation $m$.
They have non-trivial mutual braiding statistics with the braiding phase $e^{2\pi i/\mathrm{K}}$.

The BF theory \eqref{BF bulk action}
has $\mathbb{Z}_2$ charge-conjugation symmetry:
\begin{align}
	b \to -b,
	\quad
	a\to -a,
\end{align}
and our goal in this section is to
gauge this symmetry.
We will show that the resulting gauge theory has
a non-Abelian topological order.

Similar to the $(2+1)$d case, after gauging $\mathbb{Z}_2$ symmetry, 
there will be a $\mathbb{Z}_2$ charge $j$ which is a bosonic particle,
and a $\mathbb{Z}_2$ flux
which, in (3+1)d, is a vortex-line excitation.
The original excitations $e^am^b$ in the BF theory
are not $\mathbb{Z}_2$ symmetry invariant
and will be grouped together.
We will denote them simply as
$e^am^b+e^{\mathrm{K}-a}m^{\mathrm{K}-b}$.

Our approach here is to
generalize orbifolds of $(1+1)$d edge theories
discussed in Sec.\ \ref{Orbifolding 1+1 boundary theory},
and discuss orbifolds of $(2+1)$ gapless boundary theories.
In particular,
we define a set of quantities
which are analogous to the characters defined for $(1+1)$d CFTs.
These characters are constructed by applying a projection operator on the partition function,
which is equivalent to imposing a twisted boundary condition in the time direction.
The characters form a complete basis under $SL(3,\mathbb{Z})$,
the mapping class group of $T^3$,
and the $\mathcal{T}$ matrix takes a diagonal form.
By studying the characters on the boundary,
we can extract information about the non-Abelian bulk topological order.

%
%
%

\subsection{The BF surface theory}

Our starting point is thus
the boundary of the BF theory at level $\mathrm{K}$, which
can be described by the following Lagrangian density
\begin{align}
\mathcal{L}&=
\frac{\mathrm{K}}{2\pi} (\epsilon_{ij} \partial_i \zeta_j)(\partial_t \varphi)
\nonumber \\
&\quad
-\frac{1}{2\lambda_1} (\epsilon_{ij} \partial_i \zeta_j)^2
-\frac{1}{2\lambda_2}
 G^{ij} \partial_i \varphi \partial_j \varphi,
 \label{surf lag}
\end{align}
where $i,j=x,y$,
$\varphi$ is a scalar, and $\zeta_{i}$ is a one-form field
(the temporal component of $\zeta$ is gauge-fixed to zero for convenience).
We fix the coupling constant $\lambda_1$ and $\lambda_2$ according to
\begin{align}
 \frac{(2\pi)^2}{\mathrm{K}^2 \lambda_1 \lambda_2} = 1,
 \quad
 \lambda_1 = \frac{1}{\mathrm{K}},
 \quad
 \lambda_2 = \frac{(2\pi)^2}{\mathrm{K}}.
\end{align}
for convenience.
The boson field $\varphi$ is compact and satisfy
$
\varphi\equiv\varphi+2\pi.
$
Hence, physical observables are exponentials
\begin{align}
\exp [i m \varphi(t,x,y)],
\quad
m\in \mathbb{Z}.
\label{bosonic exp}
\end{align}
The winding number of $\varphi$ is quantized
in the absence of bulk quasiparticles, according to
\begin{align}
 \oint dx^i \partial_i \varphi = 2\pi N_i,
\quad
N_i\in \mathbb{Z},
\end{align}
where $i=1,2$ and $i$ is not summed on the right hand side.
On the other hand,
the gauge field $\zeta_i$ is compact, meaning that physical observables are
Wilson loops,
\begin{align}
\exp\left( i m \int_C dx^i \zeta_i(t,x,y) \right),
 \quad
 m\in \mathbb{Z},
 \label{Wilson loops}
\end{align}
where $C$ is a closed loop on $\partial\Sigma=T^2$.
(Since the different components of $\zeta_i$ commute with each other,
path-ordering is unecessary.)
The flux associated to $\zeta_i$ is quantized,
in the absence of bulk quasiparticles, according to
\begin{align}
\int dxdy\, \epsilon_{ij} \partial_i \zeta_j = 2\pi N_0
\label{compact cond for zeta}
\end{align}
where $N_0$ is an integer.

The surface theory \eqref{surf lag}
is put on a flat spacetime three-torus $T^3$,
and $G^{ij}$ represents the spatial part of
the metric. For the properties of the
the flat $T^3$, and our parameterization of
the metric, see Appendix \ref{Modular transformations on $T^3$}.
Our flat three-torus is parameterized by six real parameters,
$R_{0,1,2}$ and $\alpha,\beta,\gamma$.
For example, $R_{0,1,2}$ are the periods in $t$, $x$, and $y$ directions,
respectively.
The mapping class group of $T^3$,
${SL}(3,\mathbb{Z})$,
is generated by two transformations
which we call $U_1$ and $U_2$.
Any ${SL}(3,\mathbb{Z})$ transformation can
be written as $U^{n_1}_1 U^{n_2}_2 U^{n_3}_1\cdots$. \cite{Generators}
We further decompose
$U_1$ as
$ U_1 = U'_1M$,
where
$U_1^{\prime}$ corresponds to a $90^{\circ}$ rotation in the $\tau-x$ plane
and $M$ is a $90^{\circ}$ rotation in the $x-y$ plane. In particular,
$U'_1$ sends $\tau \to -1/\tau$
where
$
\tau= \alpha + i R_0/R_1
$.
The two transformations $U'_1$ and $U_2$ correspond respectively to
modular $S$ and $T^{-1}$ transformations in the ${\tau}-x$ plane, generating the ${SL}(2,\mathbb{Z})$ subgroup of ${SL}(3,\mathbb{Z})$ group. Combined with $M$, they generate the whole ${SL}(3,\mathbb{Z})$ group.
In the following, we denote $U^{\prime}_1 M$ by $\mathcal{S}$
and $U_2$ by $\mathcal{T}^{-1}$.

The surface theory \eqref{surf lag}
can be studied in the presence of the
following twisted boundary conditions:
\begin{align}
\varphi(t,x+2\pi R_1,y)&= \varphi(t,x,y) +
2\pi \left(
N_1 +\frac{n_1}{\mathrm{K}}
\right),
\nonumber \\
\varphi(t,x,y+2\pi R_2)&=
\varphi(t,x,y) +
2\pi
\left(
N_2 +\frac{n_2}{\mathrm{K}}
\right),
\nonumber \\
\int dxdy\, \epsilon_{ij} \partial_i \zeta_j &=
2\pi \left(
N_0 + \frac{n_0}{\mathrm{K}}
\right),
\end{align}
where $N_{0,1,2}\in \mathbb{Z}$,
and $n_{0,1,2}= 0, \ldots, \mathrm{K}-1$.
We denote the corresponding partition functions
as
\begin{align}
{Z}^{n_0n_1n_2}=Z^{n_0,n_1,n_2}_{zero}Z_{osc}
\end{align}
where $Z^{n_0,n_1,n_2}_{zero}$ and $Z_{osc}$ will be explained later. From the bulk point of view, these (twisted) boundary conditions
correspond to insertion of Wilson loop and Wilson surfaces, i.e., bulk excitations.
The bulk-boundary correspondence implies that, by studying the partition functions of
the surface theories in the presence of these boundary conditions,
in particular, their transformation law under $SL(3,\mathbb{Z})$,
one can extract properties of bulk quasiparticles.\cite{MoradiWen2015}. The details for the calculation of
surface partition functions can be found in Ref.\ \onlinecite{chen2015bulk}, where the bulk-(gapless) boundary correspondence is also discussed. Here we directly write down the partition functions for the surface. The zero mode part is

\begin{widetext}
\begin{align}
Z^{n_{0}n_{1}n_{2}}_{zero}&=
\sum_{N_{0,1,2}\in\mathbb{{Z}}}
\exp\Big\{-\frac{{\pi}\mathrm{K}^2\tau_2}{2 \mathrm{r}^2 R_2}
\tilde{N}_0^2
-2\mathrm{r}^2 \pi R_{2}\tau_{2}\left[\tilde{N}_{1}+\beta\tilde{N}_2 \right]^{2}
-\frac{{2\mathrm{r}^2\pi R_{0}R_{1}}}{R_{2}} \tilde{N}_2^2
\nonumber \\
&\qquad
+2\pi i\tau_{1}\mathrm{K}\tilde{N}_0 \left[ \tilde{N}_1 +\beta\tilde{N}_2 \right]
+2\pi i\gamma \mathrm{K}\tilde{N}_0 \tilde{N}_2 \Big\},
\label{part_bf}
\end{align}
\end{widetext}
where
$
2\mathrm{r}^2= \mathrm{K}
$,
and
we have introduced the notation
\begin{align}
\tilde{N}_{\mu}:=N_{\mu} + n_{\mu}/\mathrm{K}.
\end{align}
For the oscillator part,
\begin{align}
Z_{osc}
&=\left|\frac{1}{\eta(\tau)}\right|^2\prod_{s_2\in\mathbb{Z}^+}
\Theta_{[\beta s_2,\gamma s_2]}^{-1}\left(\tau,\frac{R_1}{R_2}s_2\right),
\end{align}
where $\Theta_{[a,b]}(\tau,m)$ is the massive theta function
  (see Appendix \ref{Theta functions}).
The total partition function for each sector is ${Z}^{n_0n_1n_2}=Z^{n_0n_1n_2}_{zero}Z_{osc}$.
The modular $\mathcal{S}$
and $\mathcal{T}$ matrices are given by
\begin{align}
\mathcal{S}_{n_i,n^{\prime}_i}&=
\frac{1}{\mathrm{K}}\delta_{n_1,n_2^{\prime}}e^{-\frac{2\pi i}{\mathrm{K}}(n^{\prime}_0n_2-n_0n_1^{\prime})},
\nonumber\\
\mathcal{T}_{n_i, n^{\prime}_i}&=
\delta_{n_0,n_0^{\prime}}\delta_{n_1,n_1^{\prime}}\delta_{n_2,n_2^{\prime}}e^{\frac{2\pi i}{\mathrm{K}}n_0n_1}.
\label{stmatrix}
\end{align}

\subsection{Gauging $\mathbb{Z}_2$ symmetry in
the surface theory}

In terms of
the surface theory \eqref{surf lag},
the $\mathbb{Z}_2$ charge conjugation symmetry
is implemented as
\begin{align}
\varphi\to-\varphi,\quad \zeta \to -\zeta.
\end{align}
We now gauge the $\mathbb{Z}_2$ symmetry.
We consider the cases of odd and even level $\mathrm{K}$ separately.

First, we include
the twisted sectors obtained by twisting
boundary conditions in $t$, $x$, and $y$ directions.
The partition functions for these sectors
are denoted by $V^{\mu,\nu,\lambda}$
where
$\mu,\nu, \lambda=0$ $(1/2)$ represents the untwisted (twisted) boundary condition,
for $t$, $x$ and $y$ directions, respectively.
The partition functions with twisted boundary condition
in $x,y,t$ directions can be readily computed and are given, respectively, by
\begin{align}
V^{0,\frac{1}{2},0}&=\left|\frac{\eta(\tau)}{\theta_4(\tau)}\right|\prod_{s_2>0}
\Theta_{[\frac{1}{2}+\beta s_2,\gamma s_2]}^{-1}\left(\tau,\frac{R_1}{R_2}s_2\right),
\nonumber \\
V^{0,0,\frac{1}{2}}&=\prod_{s_2>0}
\Theta_{[\beta s_2-\frac{1}{2},\gamma (s_2-\frac{1}{2})]}^{-1}
\left(\tau,\frac{R_1}{R_2}(s_2-\frac{1}{2})\right),
\nonumber \\
V^{\frac{1}{2},0, 0}
&=\left|\frac{\eta(\tau)}{\theta_2(\tau)}\right|\prod_{s_2>0}
\Theta_{[\beta s_2,\gamma s_2+\frac{1}{2}]}^{-1}
\left(\tau,\frac{R_1}{R_2}s_2\right),
\end{align}
($s_2$ is an integer).
All the other twisted sectors can be obtained by considering modular transformations:
\begin{align}
&
V^{[\mu],[\nu],[\lambda]}\overset{U_1^{\prime}}{\longrightarrow} V^{[\nu],[\mu],[\lambda]},
\nonumber\\
&
V^{[\mu],[\nu],[\lambda]}\overset{M}{\longrightarrow} V^{[\mu],[\lambda],[\nu]},
\nonumber\\
&
V^{[\mu],[\nu],[\lambda]}\overset{U_2}{\longrightarrow} V^{[\mu+\nu],[\nu],[\lambda]},
\end{align}
where $[a]=a\ \mbox{mod}\ \mathbb{Z}$. 
In total, there are 7 sectors with twisted boundary conditions.
(To obtain this result,
we note
several properties of
the massive theta functions $\Theta_{[a.b]}(\tau,m)$,
listed in Appendix \ref{Theta functions}.)

\begin{table}[t]
\centering
\begin{ruledtabular}
\begin{tabular}{lccc}
character $\chi$ & $d_\chi$ &
$h_\chi$
& $\mathcal{N}$\\\hline
$\chi_I=\frac{1}{2}Z_{0,0,0}+\frac{1}{2}V^{\frac{1}{2},0,0}$ & $1$ & $0$ & $1$\\
$\chi_j=\frac{1}{2}Z_{0,0,0}-\frac{1}{2}V^{\frac{1}{2},0,0}$ & $1$ & $0$ & $1$\\
$\chi_{a,b,c}=\frac{1}{2}Z_{a,b,c}+\frac{1}{2}Z_{\mathrm{K}-a,\mathrm{K}-b,\mathrm{K}-c}$ & $2$ & $\frac{ab}{\mathrm{K}}$ & $\frac{\mathrm{K}^3-1}{2}$\\
$\chi_{\sigma_x}=\frac{1}{2}V^{0,\frac{1}{2},0}+\frac{1}{2}V^{\frac{1}{2},\frac{1}{2},0}$ & $\mathrm{K}$ & $0$ & $1$\\
$\chi_{\tau_x}=\frac{1}{2}V^{0,\frac{1}{2},0}-\frac{1}{2}V^{\frac{1}{2},\frac{1}{2},0}$ & $\mathrm{K}$ & $\frac{1}{2}$ & $1$\\
$\chi_{\sigma_y}=\frac{1}{2}V^{0, 0,\frac{1}{2}}+\frac{1}{2}V^{\frac{1}{2}, 0,\frac{1}{2}}$ & $\mathrm{K}$ & $0$ & $1$\\
$\chi_{\tau_y}=\frac{1}{2}V^{0, 0,\frac{1}{2}}-\frac{1}{2}V^{\frac{1}{2}, 0,\frac{1}{2}}$ & $\mathrm{K}$ & $0$ & $1$\\
$\chi_{\sigma_{xy}}=\frac{1}{2}V^{0,\frac{1}{2},\frac{1}{2}}+\frac{1}{2}V^{\frac{1}{2},\frac{1}{2},\frac{1}{2}}$ & $\mathrm{K}$ & $0$ & $1$\\
$\chi_{\tau_{xy}}=\frac{1}{2}V^{0,\frac{1}{2},\frac{1}{2}}-\frac{1}{2}V^{\frac{1}{2},\frac{1}{2},\frac{1}{2}}$ & $\mathrm{K}$ & $\frac{1}{2}$ & $1$\\
\end{tabular}
\end{ruledtabular}
\caption{
The characters for
the single-component BF theory
after gauging the charge-conjugate $\mathbb{Z}_2$ symmetry
when $\mathrm{K}$ is odd:
The quantum dimensions $d_\chi$,
spin statistics $\theta_\chi=e^{2\pi ih_\chi}$,
and the number $\mathcal{N}$ of characters $\chi$
($h_{\chi}$ are defined only modulo $\mathbb{Z}$.)
}
\label{cc_z2_K_odd}
\end{table}

We use the twisted sectors $V^{\mu,\nu,\lambda}$ and $Z_{a,b,c}$ to build up the characters
of the gauged $(2+1)$d surface theory.
The result is listed in Table \ref{cc_z2_K_odd}.
Here, the subscript $h$ of $\chi_h$ 
denotes the type of bulk excitations. 
As mentioned previously,
the characters are constructed by inserting projection operators in the partition function.
This is similar to the construction of the minimal entangled state (MES) defined
for $(3+1)$d topological phases.\cite{MoradiWen2015, JiangMesarosRan2014}
There are $(\mathrm{K}^3+15)/2$ characters in total,
the same as the bulk ground state degeneracy on $T^3$.

As in the $(1+1)$d edge theory, 
here we have the vacuum $I$, $\mathbb{Z}_2$ boson $j$ and superselection sector $a,b,c$.  $\sigma_x$ corresponds to the bulk $\mathbb{Z}_2$ flux
excitation which leaves a twist in the $x$ direction on the boundary. The
character $\chi_{\sigma_x}$ includes partition functions with twisted boundary
condition in the $x$ direction. $\tau_x$ can be understood as combining
$\sigma_x$ with $\mathbb{Z}_2$ charge $j$ and therefore is a flux-charge
composite excitation. $\chi_{\tau_x}$ is also linear combination of partition
functions with twist boundary in $x$ direction. $\chi_{\sigma_y}$ and
$\chi_{\tau_y}$ are the characters with twisted boundary condition in $y$
direction.  $\chi_{\sigma_{xy}}$ is twisted in both the $x$ and $y$ directions.
Under the transformation $\mathcal{T}$, $\chi_{h}$ will pick up a phase $\exp 2\pi i h_{\chi}$, where $h_{\chi}$ encodes information related to $(3 + 1)$d analogue of topological spins.

We now consider the $U_1^{\prime}$ matrix which, after dimensional reduction, is the $\mathcal{S}$ matrix for a $(1+1)$d CFT.  The $U_1^{\prime}$ matrix encodes
braiding information about bulk excitations,
\begin{align}
U_1^{\prime}=\frac{1}{2\mathrm{K}}\left(\begin{smallmatrix}
1&1&2&\mathrm{K}&\mathrm{K}&0&0&0&0\\
1&1&2&-\mathrm{K}&-\mathrm{K}&0&0&0&0\\
2&2&4\cos[\frac{2\pi}{\mathrm{K}}(ab^{\prime}+ba^{\prime})]&0&0&0&0&0&0\\
\mathrm{K}&-\mathrm{K}&0&\mathrm{K}&-\mathrm{K}&0&0&0&0\\
\mathrm{K}&-\mathrm{K}&0&-\mathrm{K}&\mathrm{K}&0&0&0&0\\
0&0&0&0&0&\mathrm{K}&\mathrm{K}&\mathrm{K}&\mathrm{K}\\
0&0&0&0&0&\mathrm{K}&\mathrm{K}&-\mathrm{K}&-\mathrm{K}\\
0&0&0&0&0&\mathrm{K}&-\mathrm{K}&\mathrm{K}&-\mathrm{K}\\
0&0&0&0&0&\mathrm{K}&-\mathrm{K}&-\mathrm{K}&\mathrm{K}\\
\end{smallmatrix}\right)
\end{align}
From the first column (row) in the above matrix, we can extract the ``quantum dimension" of the bulk excitation $d_{\chi}$ shown in Table \ref{cc_z2_K_odd}.
When $\mathrm{K}=1$, the characters for the twist sectors have $d_\chi=1$. In this case, it is easy to verify that $U_1^{\prime}$ matrix and $h_{\chi}$ match up with that for
the ordinary $(3+1)$d $\mathbb{Z}_2$ gauge theory 
(the $(3+1)$d toric code model). When $\mathrm{K}>1$, the characters for the twist sectors have $d_\chi>1$, suggesting that they are non-Abelian excitations. We will discuss these non-Abelian braiding statistics later in Sec. \ref{Discussion}.

We can similarly calculate the characters when $\mathrm{K}$ is even.
Unlike in the case of $\mathrm{K}$ odd,
there are several excitations in the original BF theory that are invariant under $\mathbb{Z}_2$ symmetry operation. They will remain in the gauged topological phase and can couple with the $\mathbb{Z}_2$ boson to form composite excitations.
They will also provide species labels
for $\sigma$ and $\tau$; the characters are shown in Table \ref{cc_z2_K_even}.
There are $\mathrm{K}^3/2+60$ characters in total,
which describes the ground state degeneracy in the bulk topological phase on $T^3$.

\begin{table}[t]
\centering
\begin{ruledtabular}
\begin{tabular}{lccc}
character $\chi$ & $d_\chi$ & $h_\chi$ & $\mathcal{N}$\\\hline
$\chi_I=\frac{1}{2}Z_{0,0,0}+V^{\frac{1}{2},0,0}$ & $1$ & $0$ & $1$\\
$\chi_j=\frac{1}{2}Z_{0,0,0}-V^{\frac{1}{2},0,0}$ & $1$ & $0$ & $1$\\
$\chi_{\frac{\mathrm{K}}{2},0,0}^i=\frac{1}{2}Z_{\frac{\mathrm{K}}{2},0,0}$ & $1$ & $0$ & $2$\\
$\chi_{0,\frac{\mathrm{K}}{2},0}^i=\frac{1}{2}Z_{0,\frac{\mathrm{K}}{2},0}$ & $1$ & $0$ & $2$\\
$\chi_{\frac{\mathrm{K}}{2},\frac{\mathrm{K}}{2},0}^i=\frac{1}{2}Z_{\frac{\mathrm{K}}{2},\frac{\mathrm{K}}{2},0}$ & $1$ & $\frac{\mathrm{K}}{4}$ & $2$\\
$\chi_{0,\frac{\mathrm{K}}{2},\frac{\mathrm{K}}{2}}^i=\frac{1}{2}Z_{0,\frac{\mathrm{K}}{2},\frac{\mathrm{K}}{2}}$ & $1$ & $0$ & $2$\\
$\chi_{\frac{\mathrm{K}}{2},0,\frac{\mathrm{K}}{2}}^i=\frac{1}{2}Z_{\frac{\mathrm{K}}{2},0,\frac{\mathrm{K}}{2}}$ & $1$ & $0$ & $2$\\
$\chi_{0,0,\frac{\mathrm{K}}{2}}^i=\frac{1}{2}Z_{0,0,\frac{\mathrm{K}}{2}}$ & $1$ & $0$ & $2$\\
$\chi_{\frac{\mathrm{K}}{2},\frac{\mathrm{K}}{2},\frac{\mathrm{K}}{2}}^i=\frac{1}{2}Z_{\frac{\mathrm{K}}{2},\frac{\mathrm{K}}{2},\frac{\mathrm{K}}{2}}$ & $1$ & $\frac{\mathrm{K}}{4}$ & $2$\\
$\chi_{a,b,c}=\frac{1}{2}Z_{a,b,c}+\frac{1}{2}Z_{\mathrm{K}-a,\mathrm{K}-b,\mathrm{K}-c}$ & $2$ & $\frac{ab}{\mathrm{K}}$ & $\frac{\mathrm{K}^3-8}{2}$\\
$\chi_{\sigma_x}^m=\frac{1}{2}V^{0,\frac{1}{2},0}+\frac{1}{2}V^{\frac{1}{2},\frac{1}{2},0}$ & $\frac{\mathrm{K}}{2}$ & $0$ & $8$\\
$\chi_{\tau_x}^m=\frac{1}{2}V^{0,\frac{1}{2},0}-\frac{1}{2}V^{\frac{1}{2},\frac{1}{2},0}$ & $\frac{\mathrm{K}}{2}$ & $\frac{1}{2}$ & $8$\\
$\chi_{\sigma_y}^m=\frac{1}{2}V^{0, 0,\frac{1}{2}}+\frac{1}{2}V^{\frac{1}{2}, 0,\frac{1}{2}}$ & $\frac{\mathrm{K}}{2}$ & $0$ & $8$\\
$\chi_{\tau_y}^m=\frac{1}{2}V^{0, 0,\frac{1}{2}}-\frac{1}{2}V^{\frac{1}{2}, 0,\frac{1}{2}}$ & $\frac{\mathrm{K}}{2}$ & $0$ & $8$\\
$\chi_{\sigma_{xy}}^m=\frac{1}{2}V^{0,\frac{1}{2},\frac{1}{2}}+\frac{1}{2}V^{\frac{1}{2},\frac{1}{2},\frac{1}{2}}$ & $\frac{\mathrm{K}}{2}$ & $0$ & $8$\\
$\chi_{\tau_{xy}}^m=\frac{1}{2}V^{0,\frac{1}{2},\frac{1}{2}}-\frac{1}{2}V^{\frac{1}{2},\frac{1}{2},\frac{1}{2}}$ & $\frac{\mathrm{K}}{2}$ & $\frac{1}{2}$ & $8$\\
\end{tabular}
\end{ruledtabular}
\caption{
The characters for
the single-component BF theory
after gauging the charge-conjugate $\mathbb{Z}_2$ symmetry
when $\mathrm{K}$ is even:
The quantum dimensions $d_\chi$,
spin statistics $\theta_\chi=e^{2\pi ih_\chi}$,
and the number $\mathcal{N}$ of characters $\chi$
($h_{\chi}$ are defined only modulo $\mathbb{Z}$.)
}
\label{cc_z2_K_even}
\end{table}

The new (3+1)d topological phase which we have obtained by gauging 
the $\mathbb{Z}_2$ symmetry is the $D(\mathcal{D}_{\mathrm{K}})$ quantum double model,
where $\mathcal{D}_{\mathrm{K}}$ is the dihedral group of order $2\mathrm{K}$.
The excitations for this quantum double model are labeled by $(\mathcal{C},\rho)$,
where $\mathcal{C}$ denotes the conjugacy class and $\rho$ denotes the irreducible representation of the normalizer group for each conjugacy class.
When $\mathrm{K}$ is odd, there are in total $\frac{n+3}{2}+\frac{n-1}{2}\times n+2=\frac{n^2+7}{2}$ excitations. 
When $\mathrm{K}$ is even, there are in total $(\frac{n}{2}+3)\times 2+(\frac{n}{2}-1)\times n+ 2\times 4=\frac{n^2+28}{2}$ excitations. 
These results are consistent with the calculation for the boundary orbifold theory. 
For instance, when $\mathrm{K}=3$, $\mathcal{D}_3$ is equivalent to the symmetric group $S_3$. There are eight excitations in the $D(S_3)$ quantum double model,
which agree with the above counting.
Apart from the vacuum sector and two charge (particle) excitations, all the other five excitations are flux or flux-charge composite excitations with non-Abelian fusion rules. 
These fusion rules have been discussed in Ref.\ \onlinecite{MoradiWen2015} and will not be analyzed further here.

More generically, 
for all (3+1)d Abelian topological phases considered in this paper, 
after gauging the $\mathbb{Z}_2$ symmetry, 
the new topological phases can always be described by the quantum double model 
with group $\widetilde{\mathcal{G}}=\mathcal{G}\rtimes\mathbb{Z}_2$, 
where $\mathcal{G}$ is the original abelian group. 
This is because the $\mathbb{Z}_2$ symmetry acts on both the charge and flux excitations independently and in the same way. 
If $\mathcal{G}=\mathbb{Z}_{\mathrm{K}}$, i.e., 
the $\mathbb{Z}_{\mathrm{K}}$ gauge theory, 
the gauged model is the $D(\mathcal{D}_{\mathrm{K}})$ quantum double model.

\section{Gauging $\mathbb{Z}_2$ symmetry in
(3+1)d
$\mathbb{Z}_{\mathrm{K}}\times\mathbb{Z}_{\mathrm{K}}$ gauge theories}

In this section, we consider ordinary
$\mathbb{Z}_{\mathrm{K}}\times \mathbb{Z}_{\mathrm{K}}$
topological gauge theory
in $(3+1)$ dimensions with only particle-loop braiding statistics.
The excitations in this model can be denoted by $e_1^{a_1}m_1^{b_1}e_2^{a_2}m_2^{b_2}$, where $e_1,\ m_1,\ e_2,\ m_2$ are the fundamental excitations of this model.
This model can be described by the two-component BF theory with both components
at level $\mathrm{K}$.

The topological $\mathbb{Z}_{\mathrm{K}}\times \mathbb{Z}_{\mathrm{K}}$ gauge theory
has a $\mathbb{Z}_2$ symmetry which exchanges
\begin{align}
e_1,\ m_1\leftrightarrow e_2,\ m_2.
\end{align}
Gauging this symmetry will lead to a non-Abelian topological phase.

The surface theory of the topological
$\mathbb{Z}_{\mathrm{K}}\times \mathbb{Z}_{\mathrm{K}}$ gauge theory
can be described by taking two copies of
\eqref{surf lag}.
The surface partition function can then be written down
as
\begin{align}
W_{n_0,n_1,n_2}^{l_0,l_1,l_2}=Z_{n_0,n_1,n_2}Z_{l_0,l_1,l_2}
\end{align}
where $n_i,l_i\in\mathbb{Z}_{\mathrm{K}}$ and $i=0,1,2$.

After orbifolding the $\mathbb{Z}_2$ symmetry, the partition function with anti-periodic boundary condition in the $t$ direction is
\begin{align}
Y^{\frac{1}{2},0,0}_{n_0,n_1,n_2}=
V^{\frac{1}{2},0,0}Z^{2\mathrm{K}}_{2n_0,2n_1,2n_2}
\end{align}
where $V^{[\mu],[\nu],[\lambda]}$ is defined in the previous section and $Z^{2\mathrm{K}}_{a,b,c}$ represents the surface partition function for $\mathbb{Z}_{2\mathrm{K}}$ gauge theory model. The term $Z^{2\mathrm{K}}_{2n_i}$ is obtained by identifying $n_i$ and $l_i$ in $Z_{n_0,n_1,n_2}Z_{l_0,l_1,l_2}$, which is imposed by the $\mathbb{Z}_2$ symmetry.

The other twisted partition functions
$Y^{\mu,\nu,\lambda}_{n_0,n_1,n_2}$
can be obtained, starting from
$Y^{\frac{1}{2},0,0}_{n_0,n_1,n_2}$,
by applying modular transformations.

Schematically,
\begin{align}
Y^{\frac{1}{2},0,0}_{m_0,m_1,m_2}
&\stackrel{U'_1} {\to}
Y^{0,\frac{1}{2},0}_{n_0,n_1,n_2},
\nonumber \\
Y^{0,\frac{1}{2},0}_{m_0,m_1,m_2}
&\stackrel{U_2} {\to}
Y^{\frac{1}{2},\frac{1}{2},0}_{n_0,n_1,n_2},
\nonumber \\
Y^{0,\frac{1}{2},0}_{m_0,m_1,m_2}
&\stackrel{M}{\to}
Y^{0,0,\frac{1}{2}}_{n_0,n_1,n_2},
\nonumber \\
Y^{\frac{1}{2},\frac{1}{2},0}_{m_0,m_1,m_2}
&\stackrel{M}{\to}
Y^{\frac{1}{2},0,\frac{1}{2}}_{n_0,n_1,n_2},
\nonumber \\
Y^{\frac{1}{2},0,\frac{1}{2}}_{m_0,m_1,m_2}
&\stackrel{U'_1}{\to}
Y^{0,\frac{1}{2},\frac{1}{2}}_{n_0,n_1,n_2},
\nonumber \\
Y^{0,\frac{1}{2},\frac{1}{2}}_{m_0,m_1,m_2}
&\stackrel{U'_1}{\to}
Y^{\frac{1}{2},\frac{1}{2},\frac{1}{2}}_{n_0,n_1,n_2}.
\end{align}
The explicit form of other sectors are obtained by requiring them to be invariant (up to a phase) under $U_1^{\prime}$ transformation. 
They are given by
\begin{align}
Y^{0,\frac{1}{2},0}_{m_0,m_1,m_2}&=
V^{0,\frac{1}{2},0}\sum_{p,q}Z^{2\mathrm{K}}_{m_0+p\mathrm{K},m_1+q\mathrm{K},2m_2},
\nonumber \\
Y^{\frac{1}{2},\frac{1}{2},0}_{m_0,m_1,m_2}&=
V^{\frac{1}{2},\frac{1}{2},0}\times
\left[Z^{2\mathrm{K}}_{m_0,m_1,2m_2}\right.\nonumber\\
&+(-1)^{m_1}Z^{2\mathrm{K}}_{m_0+\mathrm{K},m_1,2m_2}
+(-1)^{m_0}Z^{2\mathrm{K}}_{m_0,m_1+\mathrm{K},2m_2}
\nonumber\\
&\left.+(-1)^{m_0+m_1+\mathrm{K}}Z^{2\mathrm{K}}_{m_0+\mathrm{K},m_1+\mathrm{K},2m_2}\right],
\nonumber \\
Y^{0,0,\frac{1}{2}}_{m_0,m_1,m_2}&=V^{0,0,\frac{1}{2}}\sum_{p,q}Z^{2\mathrm{K}}_{m_0+p\mathrm{K},2m_1, m_2+q\mathrm{K}},
\nonumber \\
Y^{\frac{1}{2},0,\frac{1}{2}}_{m_0,m_1,m_2}&=
V^{\frac{1}{2},0,\frac{1}{2}}\sum_{p,q}Z^{2\mathrm{K}}_{m_0+p\mathrm{K},2m_1, m_2+q\mathrm{K}},
\nonumber \\
Y^{0,\frac{1}{2},\frac{1}{2}}_{m_0,m_1,m_2}&=
V^{0,\frac{1}{2},\frac{1}{2}}\sum_{p,q}Z^{2\mathrm{K}}_{m_0+p\mathrm{K},2m_1, m_2+q\mathrm{K}},
\nonumber \\
Y^{\frac{1}{2},\frac{1}{2},\frac{1}{2}}_{m_0,m_1,m_2}&=
V^{\frac{1}{2},\frac{1}{2},\frac{1}{2}}\sum_{p,q}Z^{2\mathrm{K}}_{m_0+p\mathrm{K},2m_1, m_2+q\mathrm{K}},
\end{align}
where $p,q=0,1$ and $0\leq m_i< \mathrm{K}$.
$W^{l_0l_1l_2}_{n_0n_2n_2}$
and
$Y_{m_0m_1m_2}^{\nu\nu\lambda}$
can be properly combined to
construct the characters.
The result is summarized in Table \ref{cc_z2_2bf}.
It is also instructive to 
consider the dimensional reduction.
After dimensional reduction, 
the complete results for the characters on the $(1+1)$d edge are shown in Table \ref{z2_2bf_2d}.

\begin{table}[t]
\centering
\begin{ruledtabular}
\begin{tabular}{lccc}
character $\chi$ & $d_\chi$ & $h_\chi$ & $\mathcal{N}$
\\\hline
$\chi^0_{n_0,n_1,n_2}=\frac{1}{2}W_{n_0,n_1,n_2}^{n_0,n_1,n_2}+\frac{1}{2}Y_{n_0,n_1,n_2}^{\frac{1}{2},0,0}$ & $1$ & $\frac{2n_0n_1}{\mathrm{K}}$ & $\mathrm{K}^3$
\\
$\chi^1_{n_0,n_1,n_2}=\frac{1}{2}W_{n_0,n_1,n_2}^{n_0,n_1,n_2}-\frac{1}{2}Y_{n_0,n_1,n_2}^{\frac{1}{2},0,0}$ & $1$ & $\frac{2n_0n_1}{\mathrm{K}}$ & $\mathrm{K}^3$
\\
$\chi_{n_0,n_1,n_2}^{l_0,l_1,l_2}=\frac{1}{2}W_{n_0,n_1,n_2}^{l_0,l_1,l_2}+
\frac{1}{2}W_{l_0,l_1,l_2}^{n_0,n_1,n_2}$ & $2$ & $\frac{n_0n_1+l_0l_1}{\mathrm{K}}$ & $\frac{\mathrm{K}^6-\mathrm{K}^3}{2}$
\\
$\chi_{\sigma_x}^{m_0,m_1,m_2}=\frac{1}{4}Y^{0,\frac{1}{2},0}_{m_i}+\frac{1}{4}Y^{\frac{1}{2},\frac{1}{2},0}_{m_i}$ & $\mathrm{K}$ & $\frac{m_0m_1}{2\mathrm{K}}$ & $\mathrm{K}^3$\\
$\chi_{\tau_x}^{m_0,m_1,m_2}=\frac{1}{4}Y^{0,\frac{1}{2},0}_{m_i}-\frac{1}{4}Y^{\frac{1}{2},\frac{1}{2},0}_{m_i}$ & $\mathrm{K}$ & $\frac{\mathrm{K}+m_0m_1}{2\mathrm{K}}$ & $\mathrm{K}^3$\\
$\chi_{\sigma_y}^{m_0,m_1,m_2}=\frac{1}{4}Y^{0, 0,\frac{1}{2}}_{m_i}+\frac{1}{4}Y^{\frac{1}{2}, 0,\frac{1}{2}}_{m_i}$ & $\mathrm{K}$ & $\frac{m_0m_1}{\mathrm{K}}$ & $\mathrm{K}^3$\\
$\chi_{\tau_y}^{m_0,m_1,m_2}=\frac{1}{4}Y^{0, 0,\frac{1}{2}}_{m_i}-\frac{1}{4}Y^{\frac{1}{2}, 0,\frac{1}{2}}_{m_i}$ & $\mathrm{K}$ & $\frac{m_0m_1}{\mathrm{K}}$ & $\mathrm{K}^3$\\
$\chi_{\sigma_{xy}}^{m_0,m_1,m_2}=\frac{1}{4}Y^{0,\frac{1}{2},\frac{1}{2}}_{m_i}+\frac{1}{4}Y^{\frac{1}{2},\frac{1}{2},\frac{1}{2}}_{m_i}$ & $\mathrm{K}$ & $\frac{m_0m_1}{\mathrm{K}}$ & $\mathrm{K}^3$\\
$\chi_{\tau_{xy}}^{m_0,m_1,m_2}=\frac{1}{4}Y^{0,\frac{1}{2},\frac{1}{2}}_{m_i}-\frac{1}{4}Y^{\frac{1}{2},\frac{1}{2},\frac{1}{2}}_{m_i}$ & $\mathrm{K}$ & $\frac{\mathrm{K}+2m_0m_1}{2\mathrm{K}}$ & $\mathrm{K}^3$\\
\end{tabular}
\end{ruledtabular}
\caption{
The characters for
the
topological
$\mathbb{Z}_{\mathrm{K}}\times \mathbb{Z}_{\mathrm{K}}$
theory in $(3+1)$d
after gauging the $\mathbb{Z}_2$ symmetry:
The quantum dimensions $d_\chi$,
spin statistics $\theta_\chi=e^{2\pi ih_\chi}$,
and the number $\mathcal{N}$ of characters $\chi$
($h_{\chi}$ are defined only modulo $\mathbb{Z}$.)
}
\label{cc_z2_2bf}
\end{table}

\begin{table}[t]
\centering
\begin{ruledtabular}
\begin{tabular}{lccc}
character $\chi$ & $d_\chi$ & $h_\chi$ & $\mathcal{N}$\\\hline
$\chi^0_{n_0,n_1}=\frac{1}{2}W_{n_0,n_1}^{n_0,n_1}+\frac{1}{2}Y_{n_0,n_1}^{\frac{1}{2},0}$ & $1$ & $\frac{2n_0n_1}{\mathrm{K}}$ & $\mathrm{K}^2$\\
$\chi^1_{n_0,n_1}=\frac{1}{2}W_{n_0,n_1}^{n_0,n_1}-\frac{1}{2}Y_{n_0,n_1}^{\frac{1}{2},0}$ & $1$ & $\frac{2n_0n_1}{\mathrm{K}}$ & $\mathrm{K}^2$\\
$\chi_{n_0,n_1}^{l_0,l_1,l_2}=\frac{1}{2}\left(W_{n_0,n_1}^{l_0,l_1}+W_{l_0,l_1}^{n_0,n_1}\right)$ & $2$ & $\frac{n_0n_1+l_0l_1}{\mathrm{K}}$ & $\frac{\mathrm{K}^4-\mathrm{K}^2}{2}$\\
$\chi_{\sigma}^{m_0,m_1}=\frac{1}{4}Y^{0,\frac{1}{2}}_{m_i}+\frac{1}{4}Y^{\frac{1}{2},\frac{1}{2}}_{m_i}$ & $\mathrm{\mathrm{K}}$ & $\frac{m_0m_1}{2\mathrm{K}}$ & $\mathrm{K}^2$\\
$\chi_{\tau}^{m_0,m_1}=\frac{1}{4}Y^{0,\frac{1}{2}}_{m_i}-\frac{1}{4}Y^{\frac{1}{2},\frac{1}{2}}_{m_i}$ & $\mathrm{\mathrm{K}}$ & $\frac{1}{2}+\frac{m_0m_1}{2\mathrm{K}}$ & $\mathrm{K}^2$\\
\end{tabular}
\end{ruledtabular}
\caption{
	The quantum dimensions $d_\chi$, 
	spin statistics $\theta_\chi=e^{2\pi ih_\chi}$ and number $\mathcal{N}$ of characters $\chi$ 
	from orbifolding $\mathbb{Z}_2$ symmetry of boundary of the $\mathbb{Z}_{\mathrm{K}}\times \mathbb{Z}_{\mathrm{K}}$ gauge theory in $(2+1)$d. This table matches up with Table \ref{cc_z2_2bf} after dimensional reduction.}
\label{z2_2bf_2d}
\end{table}

Finally, by noting the transformation properties
of the characters under $U_1^{\prime}$
listed in Appendix \ref{transformation laws},
we read off the $U_1^{\prime}$ matrix:
\begin{widetext}
\begin{align}
U_1^{\prime}=\frac{1}{\mathcal{D}}\left(\begin{smallmatrix}
e^{\frac{2\pi i}{\mathrm{K}}(2n_0n_1^{\prime}+2n_1n_0^{\prime})}
&e^{\frac{2\pi i}{\mathrm{K}}(2n_0n_1^{\prime}+2n_1n_0^{\prime})}
&2e^{\frac{2\pi i}{\mathrm{K}}(n_0(n_1^{\prime}+l_1^{\prime})+n_1(n_0^{\prime}+l_0^{\prime}))}
&\mathrm{K}e^{\frac{2\pi i}{\mathrm{K}}(n_0m_1^{\prime}+n_1m_0^{\prime})}
&\mathrm{K}e^{\frac{2\pi i}{\mathrm{K}}(n_0m_1^{\prime}+n_1m_0^{\prime})}\\  
e^{\frac{2\pi i}{\mathrm{K}}(2n_0n_1^{\prime}+2n_1n_0^{\prime})}
&e^{\frac{2\pi i}{\mathrm{K}}(2n_0n_1^{\prime}+2n_1n_0^{\prime})}
&2e^{\frac{2\pi i}{\mathrm{K}}(n_0(n_1^{\prime}+l_1^{\prime})+n_1(n_0^{\prime}+l_0^{\prime}))}
&-\mathrm{K}e^{\frac{2\pi i}{\mathrm{K}}(n_0m_1^{\prime}+n_1m_0^{\prime})}
&-\mathrm{K}e^{\frac{2\pi i}{\mathrm{K}}(n_0m_1^{\prime}+n_1m_0^{\prime})}\\  
2e^{\frac{2\pi i}{\mathrm{K}}(n_0^{\prime}(n_1+l_1)+n_1^{\prime}(n_0+l_0))}
&2e^{\frac{2\pi i}{\mathrm{K}}(n_0^{\prime}(n_1+l_1)+n_1^{\prime}(n_0+l_0))}
&2e^{\frac{2\pi i}{\mathrm{K}}(n_0n_1^{\prime}+n_1n_0^{\prime}+l_0l_1^{\prime}+l_1l_0^{\prime})}
&0&0\\             
\mathrm{K}e^{\frac{2\pi i}{\mathrm{K}}(n^{\prime}_0m_1+n_1^{\prime}m_0)}
&-\mathrm{K}e^{\frac{2\pi i}{\mathrm{K}}(n^{\prime}_0m_1+n_1^{\prime}m_0)}
&0
&\frac{\mathrm{\mathrm{K}}}{2}Pe^{\frac{\pi i}{\mathrm{K}}(m_0m_1^{\prime}+m_1m_0^{\prime})}
&-\frac{\mathrm{\mathrm{K}}}{2}Pe^{\frac{\pi i}{\mathrm{K}}(m_0m_1^{\prime}+m_1m_0^{\prime})}\\   
\mathrm{K}e^{\frac{2\pi i}{\mathrm{K}}(n^{\prime}_0m_1+n_1^{\prime}m_0)}
&-\mathrm{K}e^{\frac{2\pi i}{\mathrm{K}}(n^{\prime}_0m_1+n_1^{\prime}m_0)}
&0
&-\frac{\mathrm{\mathrm{K}}}{2}Pe^{\frac{\pi i}{\mathrm{K}}(m_0m_1^{\prime}+m_1m_0^{\prime})}
&\frac{\mathrm{\mathrm{K}}}{2}P\mathrm{K}e^{\frac{\pi i}{\mathrm{K}}(m_0m_1^{\prime}+m_1m_0^{\prime})}
\end{smallmatrix}\right)
\end{align}
where
$\mathcal{D}=2\mathrm{K}^2$
and $P=
\left[
1+(-1)^{m_1+m_1^{\prime}}+(-1)^{m_0+m_0^{\prime}}(1+(-1)^{m_1+m_1^{\prime}+\mathrm{K}}) \right]$.
\end{widetext}

\subsection{Orbifolding phases with non-trivial 
three-loop braiding statistics}

For the $\mathbb{Z}_{\mathrm{K}}\times \mathbb{Z}_{\mathrm{K}}$ twisted gauge theory 
with non-trivial three loop braiding statistics
\cite{WangLevin2014,JiangMesarosRan2014, WangWen2015},  
we can also construct the surface partition function and calculate $\mathcal{S}$ and $\mathcal{T}$ matrices. 
\cite{chen2015bulk} 
In this case, the quantum numbers $n_0$ and $l_0$ are shifted according to
\begin{align}
 n_0 &\to \widetilde{n}_0=n_0+(l_1 n_2-l_2 n_1)/\mathrm{K},
 \nonumber\\
l_0 & \to \widetilde{l}_0=l_0+(n_1 l_2-n_2 l_1)/\mathrm{K}. 
\end{align}
This model also has the $\mathbb{Z}_2$ exchange symmetry, which switches $n_i$ and $l_i$. Therefore we can gauge the $\mathbb{Z}_2$ symmetry. After orbifolding this symmetry on the surface, for the characters $\chi^0_{n_0,n_1,n_2}$ with $n_i=l_i$, they are still the same as for the untwisted $\mathbb{Z}_{\mathrm{K}}\times \mathbb{Z}_{\mathrm{K}}$ gauge theory. The character $\chi_{n_i}^{l_i}$ is modified slightly: the quantum numbers $n_0, l_0$ replaced by $\widetilde{n}_0$ and $\widetilde{l}_0$. The characters for twisted sectors are still the same as before. The complete result is shown in Table \ref{three_loop_z2}. The $U_1^{\prime}$ matrix is very similar to the ordinary $\mathbb{Z}_{\mathrm{K}}\times \mathbb{Z}_{\mathrm{K}}$ gauge theory and we will not discuss it here.

\begin{table}[t]
\centering
\begin{ruledtabular}
\begin{tabular}{lccc}
character $\chi$ & $d_\chi$ & $h_\chi$ & $\mathcal{N}$\\\hline
$\chi^0_{n_0,n_1,n_2}=\frac{1}{2}W_{n_0,n_1,n_2}^{n_0,n_1,n_2}+\frac{1}{2}Y_{n_0,n_1,n_2}^{\frac{1}{2},0,0}$ & $1$ & $\frac{2n_0n_1}{\mathrm{K}}$ & $\mathrm{K}^3$
\\
$\chi^1_{n_0,n_1,n_2}=\frac{1}{2}W_{n_0,n_1,n_2}^{n_0,n_1,n_2}-\frac{1}{2}Y_{n_0,n_1,n_2}^{\frac{1}{2},0,0}$ & $1$ & $\frac{2n_0n_1}{\mathrm{K}}$ & $\mathrm{K}^3$
\\
$\chi_{n_0,n_1,n_2}^{l_0,l_1,l_2}=\frac{1}{2}\left(Z_{\widetilde{n}_0,n_1,n_2}^{\widetilde{l}_0,l_1,l_2}+Z_{\widetilde{l}_0,l_1,l_2}^{\widetilde{n}_0,n_1,n_2}\right)$ & $2$ & $\frac{\widetilde{n}_0n_1+\widetilde{l}_0l_1}{\mathrm{K}}$ & $\frac{\mathrm{K}^6-\mathrm{K}^3}{2}$
\\
$\chi_{\sigma_x}^{m_0,m_1,m_2}=\frac{1}{4}Y^{0,\frac{1}{2},0}_{m_i}+\frac{1}{4}Y^{\frac{1}{2},\frac{1}{2},0}_{m_i}$ & $\mathrm{K}$ & $\frac{m_0m_1}{2\mathrm{K}}$ & $\mathrm{K}^3$
\\
$\chi_{\tau_x}^{m_0,m_1,m_2}=\frac{1}{4}Y^{0,\frac{1}{2},0}_{m_i}-\frac{1}{4}Y^{\frac{1}{2},\frac{1}{2},0}_{m_i}$ & $\mathrm{K}$ & $\frac{1}{2}+\frac{m_0m_1}{2\mathrm{K}}$ & $\mathrm{K}^3$
\\
$\chi_{\sigma_y}^{m_0,m_1,m_2}=\frac{1}{4}Y^{0, 0,\frac{1}{2}}+\frac{1}{4}Y^{\frac{1}{2}, 0,\frac{1}{2}}$ & $\mathrm{K}$ & $\frac{m_0m_1}{\mathrm{K}}$ & $\mathrm{K}^3$
\\
$\chi_{\tau_y}^{m_0,m_1,m_2}=\frac{1}{4}Y^{0, 0,\frac{1}{2}}_{m_i}-\frac{1}{4}Y^{\frac{1}{2}, 0,\frac{1}{2}}_{m_i}$ & $\mathrm{K}$ & $\frac{m_0m_1}{\mathrm{K}}$ & $\mathrm{K}^3$\\
$\chi_{\sigma_{xy}}^{m_0,m_1,m_2}=\frac{1}{4}Y^{0,\frac{1}{2},\frac{1}{2}}_{m_i}+\frac{1}{4}Y^{\frac{1}{2},\frac{1}{2},\frac{1}{2}}_{m_i}$ & $\mathrm{K}$ & $\frac{m_0m_1}{\mathrm{K}}$ & $\mathrm{K}^3$\\
$\chi_{\tau_{xy}}^{m_0,m_1,m_2}=\frac{1}{4}Y^{0,\frac{1}{2},\frac{1}{2}}_{m_i}-\frac{1}{4}Y^{\frac{1}{2},\frac{1}{2},\frac{1}{2}}_{m_i}$ & $\mathrm{K}$ & $\frac{1}{2}+\frac{m_0m_1}{\mathrm{K}}$ & $\mathrm{K}^3$\\
\end{tabular}
\end{ruledtabular}
\caption{The quantum dimensions $d_\chi$,  spin statistics $\theta_\chi=e^{2\pi ih_\chi}$ and number $\mathcal{N}$ of characters $\chi$ from orbifolding $\mathbb{Z}_2$ symmetry of surface of $\mathbb{Z}_{\mathrm{K}}\times \mathbb{Z}_{\mathrm{K}}$ twisted gauge theory in $(3+1)$d. $\widetilde{n}_0=n_0+(l_1 n_2-l_2 n_1)/\mathrm{K}$ and $\widetilde{l}_0=l_0+(n_1 l_2-n_2 l_1)/\mathrm{K}$.}
\label{three_loop_z2}
\end{table}

\section{Physics in the bulk}
\label{Discussion}

In this section, we will study the bulk physics and discuss the non-Abelian braiding statistics
of loop excitations.

\paragraph{Twist defects in (2+1)-dimensional topological phases}

\begin{figure}[bt]
\centering
\includegraphics[scale=.4]{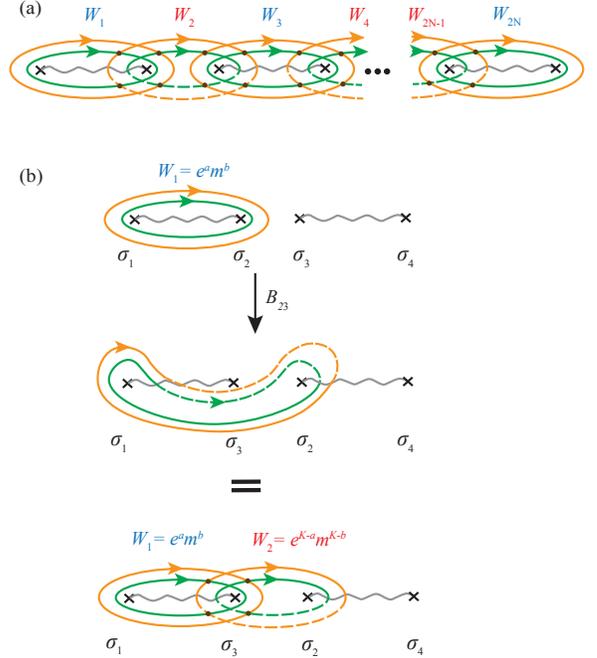}
\caption{
(a) The Wilson loop operator defined on a pair of adjacent twist defects. The green loop is for the $e$-loop and orange loop is for $m$-loop. 
The solid orange and green lines denote the Wilson loop $e^{a}m^b$ and the dotted lines represent $e^{\mathrm{K}-m}m^{\mathrm{K}-b}$.
(b) The braiding process between $\sigma_2$ and $\sigma_3$.
}
\label{fig:braid_2d}
\end{figure}

Before we discuss the $(3+1)$d case,
it is instructive to review briefly the $(2+1)$d case.\cite{Kitaev06, Bombin2010, KitaevKong12, YouWen, BarkeshliQi, BarkeshliJianQi, Teo2014}
For the $D(\mathbb{Z}_{\mathrm{K}})$ quantum double model,
we can introduce a twofold twist defect which exchanges Abelian excitations
$e^am^b$ and $e^{\mathrm{K}-a}m^{\mathrm {K}-b}$.
The twist defect is a point-like defect 
with a branch cut emanating from it.
In Fig.\ \ref{fig:braid_2d}, we depict several pairs of twist defects, with branch cuts extending between them.
When an anyon $e^am^b$ ($e^{\mathrm{K}-a}m^{\mathrm{K}-b}$)
is dragged around the twist defect, it is transformed to 
$e^{\mathrm{K}-a}m^{\mathrm{K}-b}$ ($e^am^b$) when it passes through the branch cut. 
Once the $\mathbb{Z}_2$ charge-conjugation symmetry is gauged,
twist defects, where were treated above as non-dynamical objects,  
are deconfined $\mathbb{Z}_2$ flux excitations. \cite{Barkeshli_gauging_2014, Teo2015,Tarantino2016}
These $\mathbb{Z}_2$ fluxes can leave twists on the boundary, which correspond to the twist fields $\sigma$ or $\tau$ on the boundary discussed in the previous section.
As we will see below, 
these $\mathbb{Z}_2$ fluxes are non-Abelian excitations.

The non-Abelian braiding statistics of these twist defects (before gauging)
can be studied by calculating the fundamental unitary exchange operations,
called B-symbols.\cite{BarkeshliQi, BarkeshliJianQi, Teo2014}
Each B-move represents a counter-clockwise permutation of a pair of adjacent defects,
which can result in a transformation of different ground states. 
The B-operations can be generated by a sequence of 
F and R-moves,
and evaluated exactly once we specify the splitting space of the twist defects.
\cite{Teo2014}
Here we show that twist defects are non-Abelian objects 
by directly deforming Wilson loop operators around a pair of twist defects 
shown in Fig.\ \ref{fig:braid_2d}.
Let us consider 
a system with $2N$ twist defects aligned on a line. 
The Hilbert space can be characterized by the non-contractible Wilson loop operators around the neighboring twist defects (Fig.\ \ref{fig:braid_2d} (a)).
The Wilson loop operators are represented as the powers of fundamental 
$e$-loop and $m$-loop operators. 
It is important to note that 
the neighboring Wilson loop operators do not commute with each other due to the intersections (highlighted by brown dots).
Therefore, the Hilbert space can be 
spanned
by the eigenstates of Wilson loops 
$\{W_{1},W_3,\ldots, W_{2N-1}\}$ 
where 
$W_{2j-1}$ can be either $e_{2j-1}$ or $m_{2j-1}$:
\begin{align}
&e_{2j-1}|n_1,m_1,\ldots, n_j,m_j,\ldots, n_N,m_N\rangle
\nonumber\\
&\quad
=e^{\frac{4\pi im_j}{\mathrm{K}}}|n_1,m_1,\ldots, n_j,m_j,\ldots, n_N,m_N\rangle,\nonumber\\
&m_{2j-1}|n_1,m_1,\ldots, n_j,m_j,\ldots, n_N,m_N\rangle
\nonumber\\
&\quad
=e^{\frac{-4\pi in_j}{\mathrm{K}}}|n_1,m_1,\ldots, n_j,m_j,\ldots, n_N,m_N\rangle.
\end{align}
On the other hand, 
Wilson loops
$\{W_{2},W_4,\ldots, W_{2N}\}$ 
act on the basis states as
\begin{align}
&e_{2j}|n_1,m_1,\ldots, n_j,m_j,\ldots, n_N,m_N\rangle
\nonumber\\
&\quad
=|n_1,m_1,\ldots, n_j+1,m_j,\ldots, n_N,m_N\rangle,
\nonumber\\
&m_{2j}|n_1,m_1,\ldots, n_j,m_j,\ldots, n_N,m_N\rangle
\nonumber\\
&\quad
=|n_1,m_1,\ldots, n_j,m_j+1,\ldots, n_N,m_N\rangle.
\end{align}

The non-commutative algebra between neighboring Wilson loop operators
leads to non-Abelian braiding of twist defects.
\cite{BarkeshliJianQi, You_Jian_Wen}
In Fig.\ \ref{fig:braid_2d} (b), we consider four twist defects with $W_{1}=e^am^b, W_{2}=\mathbb{I}$. If we  exchange the twist defects $\sigma_2$ and $\sigma_3$, $W_{1}$ will deform to $W_1W_2^{\prime}$ up to some phase with $W_2^{\prime}=e^am^b$ and the original state $|0,0\rangle$ changes to $|a,b\rangle$.

\paragraph{Twist defects in (3+1)-dimensional topological phases}

\begin{figure}[bt]
\centering
\includegraphics[scale=.4]{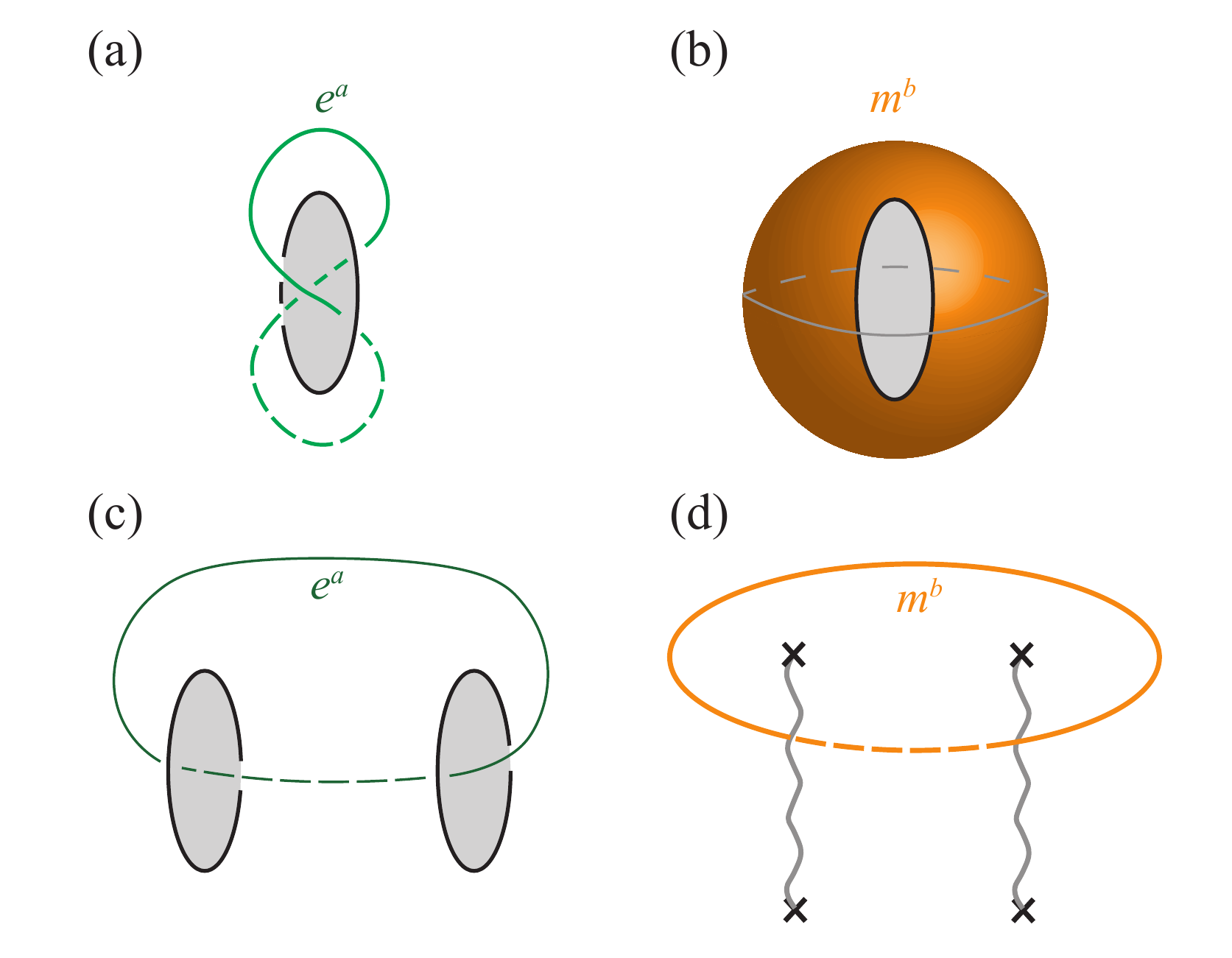}
\caption{
Wilson loop or surface operators in the presence of a twist defect. 
(a): A Wilson loop operator threading 
through the branch sheet of a twist defect. 
A twist defect is a loop (line)-like object with a branch sheet 
(represented by a shaded region). The Wilson loop must intersect the branch surface twice in
order to close: the charge $e^a$ is conjugated to $e^{K-a}$ at the first intersection
and then back to $e^a$ at the second intersection.
(b): A Wilson surface (sphere) operator wrapping around a twist defect loop. 
The surface operator cannot shrink to nothing since there is a loop living inside.
(c): A Wilson loop operator threading through the branch sheets of 
two twist defects.  
(d): A Wilson surface operator threading though the branch sheets of twist defects. Here, 
for pictorial simplicity,
the dimensionally-reduced
configuration is shown.
}
\label{fig:Wilson_3d}
\end{figure}

Similarly, we can understand the non-Abelian braiding of twist defects in three spatial dimensions. 
As shown in Fig.~\ref{fig:Wilson_3d},
the twist defect has a loop configuration in three spatial dimensions
and has a branch sheet attached to it. 
Unlike in $(2+1)$d, this defect loop does not need to pair up with another twist defect
since it is equivalent to a pair of extended defect lines that wrap around a non-contractible circle on $T^3$.  
For any excitation $e^am^b$, as it goes through the branch sheet, it will become $e^{\mathrm{K}-a}m^{\mathrm{K}-b}$.

For a system with a finite number of twist defects,
the Hilbert space is labeled by the non-contractible Wilson loop and surface operators as shown in Fig.~\ref{fig:Wilson_3d}.
In Figs.~\ref{fig:Wilson_3d} (a) and (b), 
we show the Wilson loop and surface operators defined 
in the presence of a single defect loop,
while in (c) and (d), the Wilson operators are defined 
for a pair of loops. 
By counting these Wilson operators, 
we find that the quantum dimension of a defect loop is $\mathrm{K}^2$. 
Therefore the extended defect line has quantum dimension $\mathrm{K}$ and matches up with that for the twist field on the boundary theory in Table \ref{cc_z2_K_odd}.

\begin{figure}[t]
\centering
\includegraphics[scale=.4]{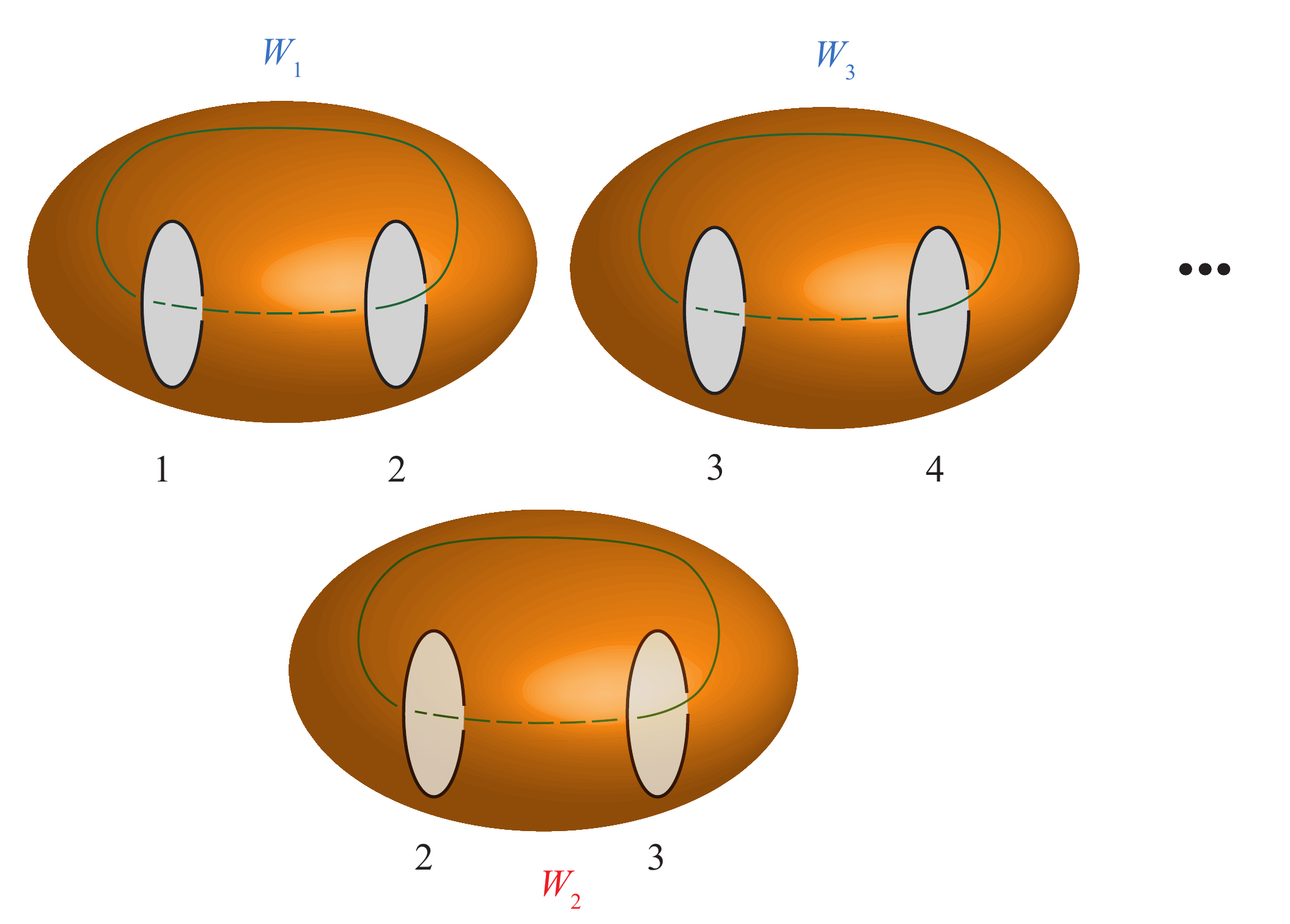}
\caption{
A subset of Wilson loop or Wilson surface operators defined 
in the presence of twist defects.
}
\label{fig:hilbert_space_3d}
\end{figure}

\begin{figure}[t]
\centering
\includegraphics[scale=.4]{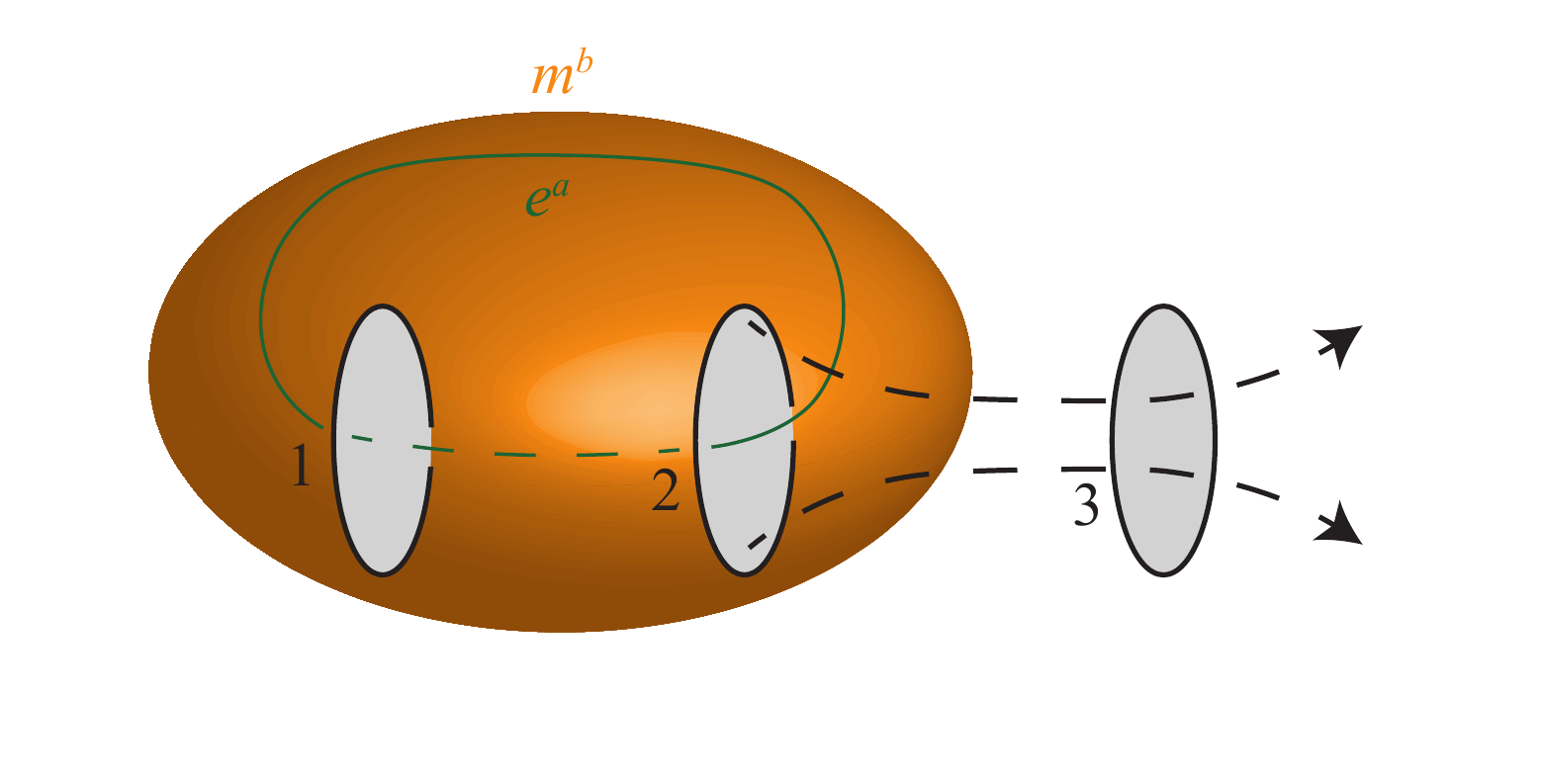}
\caption{
Braiding process between loop $2$ and loop $3$.
}
\label{fig:braid_3d}
\end{figure}

\begin{figure}[t]
\centering
\includegraphics[scale=.4]{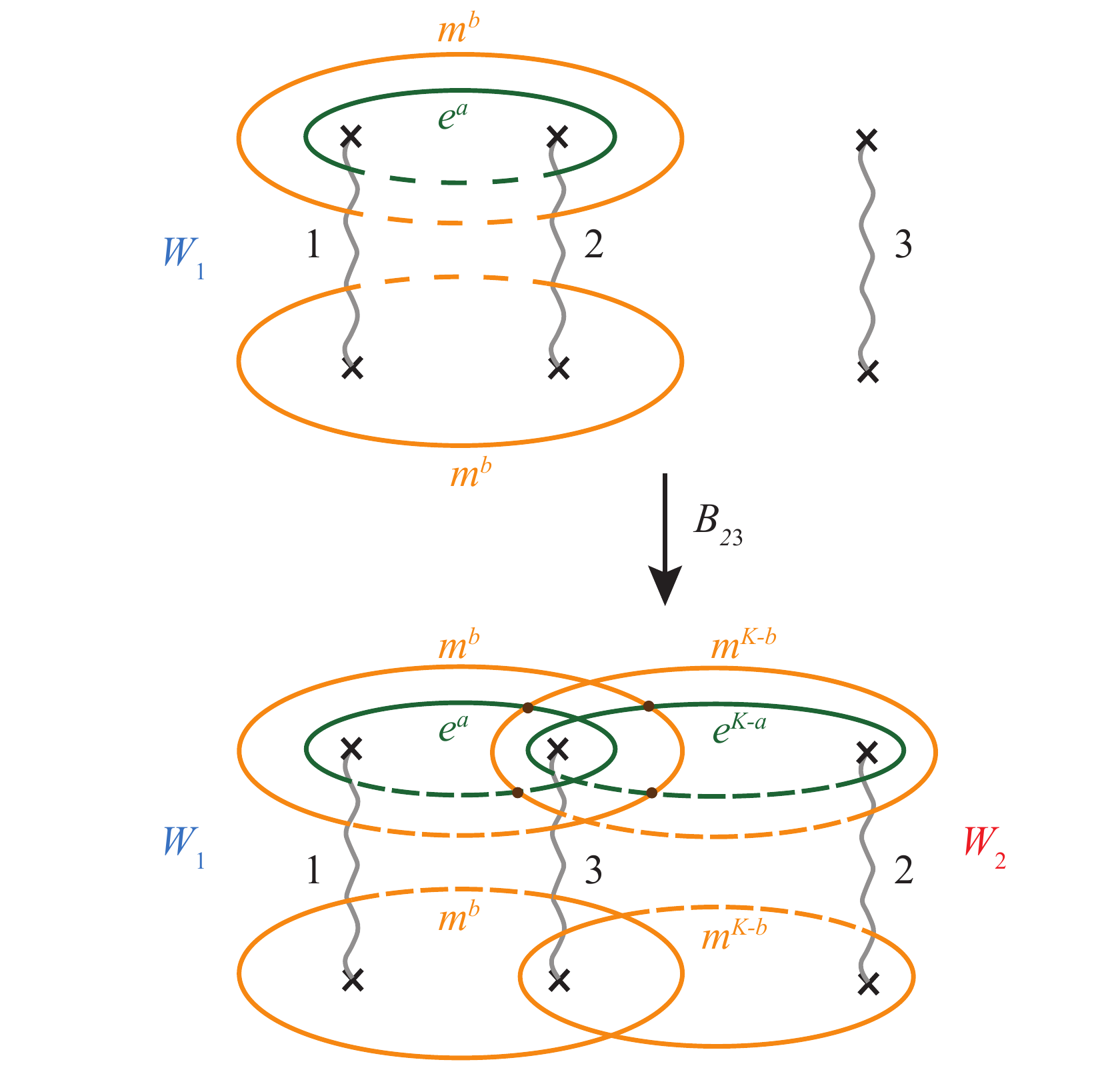}
\caption{
Braiding process between loop $2$ and loop $3$ after dimensional reduction.
}
\label{fig:braid_red_2d}
\end{figure}

Here we shall use (b) and (c) to construct a subspace of the total Hilbert space and show that
these loop excitations have non-Abelian braiding statistics.
As shown in Fig.~\ref{fig:hilbert_space_3d},
the Wilson operators $W_{2i-1}$ and $W_{2j}$ are defined on a pair of defect loops.
The adjacent Wilson operators do not commute with each other.
As in the 2d case,
the Hilbert space can be generated by acting with $\{W_{2j}\}$ operators on basis states in which $\{W_{2j-1}\}$
is diagonal. The braiding process of loop $2$ and $3$ is defined in Fig.~\ref{fig:braid_3d}.
This exchange process can be better understood 
if we look at the dimensionally-reduced version in Fig.~\ref{fig:braid_red_2d}.
$W_{1}$ under this process deforms into $W_{1}W_2$,
suggesting that a defect loop is a non-Abelian object.
Once $\mathbb{Z}_2$ symmetry is fully gauged,
the defect loops will be the intrinsic non-Abelian $\mathbb{Z}_2$ flux excitations.
These are loops in $D(\mathcal{D}_{\mathrm{K}})$ that carry
flux equal to the conjugacy class of reflections in the dihedral group $\mathcal{D}_{\mathrm{K}}$.
They fuse non-trivially because the composition of two reflections can be either the identity
or a rotation; braiding transforms the system within the state space of these different fusion outcomes. 
For the $\mathbb{Z}_\mathrm{K}\times\mathbb{Z}_\mathrm{K}$ gauge theory
(both with and without non-trivial three-loop braiding),
using similar method, we can also show that the 
twist defect loop/$\mathbb{Z}_2$ flux excitations are non-Abelian objects.

\section{Conclusion}
\label{conclusion}

In this paper, we gauge 
the $\mathbb{Z}_2$ symmetry in various Abelian topological phases in $(3+1)$ dimensions. 
By making use of the bulk-boundary correspondence, 
we discuss the orbifold theory on the $(2+1)$d surface state.
We calculate the partition function on the $(2+1)$d torus
with twisted boundary conditions and group them into characters. 
We further study how the characters transform under modular 
the $\mathcal{S}$ and $\mathcal{T}$ transformations which characterize
the braiding statistics of particle and loops excitations in the bulk.
Based on the topological data obtained on the boundary, we further analyze the defect loops/$\mathbb{Z}_2$ flux excitations in the bulk and use the Wilson loop algebra to show that these loop defects are non-Abelian.

Recently, it was shown that Abelian topological phases in $(3+1)$ dimensions,
such as the $\mathbb{Z}_\mathrm{K}$ and
$\mathbb{Z}_\mathrm{K}\times\mathbb{Z}_\mathrm{K}$ gauge theories discussed here,
have flux line excitations carrying Cheshire charge, topological charge that cannot be localized to a point
on the flux line or measured locally \cite{Else2017}. Since many properties can be deduced from those
of the parent Abelian theory, the gauging procedure discussed here may be
an entry point for exploring the properties of Cheshire charge in non-Abelian topological phases in $(3+1)$ dimensions.

All the $(3+1)$d topological phases in this paper,
obtained by gauging the $\mathbb{Z}_2$ symmetry, 
can be described by quantum double models with a non-Abelian group
and their dimensional reductions are $(2+1)$d quantum double models. 
In the future, it would be interesting to explore non-Abelian topological phases 
in $(3+1)$d that go beyond quantum double models.

\acknowledgements
We thank Michael Levin and Jeffrey Teo for useful discussions.
We also thank KITP program “Symmetry,
Topology, and Quantum Phases of Matter: From Tensor
Networks to Physical Realizations”. 
This work is supported in part by the NSF under Grant No.\ DMR-1455296 and 
No. NSF PHY-1125915. XC was supported by a postdoctoral fellowship from the Gordon and Betty Moore Foundation,
under the EPiQS initiative, Grant GBMF4304, at the Kavli Institute for Theoretical Physics.  

\appendix

\begin{widetext}

\section{Theta functions}
\label{Theta functions}

The Dedekind eta function
$\eta(\tau)$ is
defined by
\begin{align}
\eta(\tau):=e^{\frac{\pi i\tau}{12}}\prod_{n=1}^{\infty}(1-q^n),
\quad
q:=e^{2\pi i\tau}.
\end{align}

The massive theta function
$\Theta_{[a,b]}(\tau,m)$
is defined by
\begin{align}
\Theta_{[a,b]}(\tau,m)&\equiv e^{4\pi \tau_2\Delta(m,a)}
 \prod_{n\in\mathbb{Z}}\left|1-e^{-2\pi \tau_{2}\sqrt{m^2+(n+a)^2}+2\pi i\tau_{1}(n+a)+2\pi i b}\right|^2
\end{align}
where
\begin{align}
\Delta(m,a)\equiv\frac{1}{2}\sum_{n\in\mathbb{Z}}\sqrt{m^2+(n+a)^2}-\frac{1}{2}\int_{-\infty}^{\infty}dk(m^2+k^2)^{1/2}
\end{align}
The massive theta functions $\Theta_{[a.b]}(\tau,m)$
satisfy
\begin{align}
&
\Theta_{[a,b]}(\tau,m)=\Theta_{[-a,-b]}(\tau,m)=\Theta_{[a+p,b+q]}(\tau,m),
\nonumber\\
&
\Theta_{[a,b]}(\tau+1,m)=\Theta_{[a,b+a]}(\tau,m),
\nonumber\\
&
\Theta_{[a,b]}(-1/\tau,m|\tau|)=\Theta_{[b,-a]}(\tau,m), 
\end{align}
where $p,q\in\mathbb{Z}$.

\section{Modular transformations on $T^3$}
\label{Modular transformations on $T^3$}

In this appendix, we collect some necessary
ingredients relating to a flat three-torus $T^3$ and its mapping class group.\cite{Hsieh2015}
A flat three-torus is parameterized by six real parameters,
$R_{0,1,2}$ and $\alpha,\beta,\gamma$.
For a flat three-torus $T^3$, the dreibein can be factorized as
\begin{align}
 {e^A}_{\mu}
 &=
 \left(
 \begin{array}{ccc}
 R_0 & 0 & 0 \\
 0 & R_1 & 0 \\
 0 & 0 & R_2
 \end{array}
 \right)
 \left(
  \begin{array}{ccc}
 1 & 0 & 0 \\
 -\alpha & 1 & 0 \\
 -\gamma & -\beta & 1
 \end{array}
 \right)
 =
 \left(
  \begin{array}{ccc}
 R_0 & 0 & 0 \\
 -\alpha R_1 & R_1 & 0 \\
 -\gamma R_2 & -\beta R_2 & R_2
 \end{array}
 \right),
\end{align}
and its inverse is given by
\begin{align}
 {e^{\star}_A}^{\mu}=
  \left(
 \begin{array}{ccc}
 \frac{1}{R_0} & \frac{\alpha}{R_0} & \frac{\alpha\beta+\gamma}{R_0} \\
 0 & \frac{1}{R_1} & \frac{\beta}{R_1} \\
 0 & 0 & \frac{1}{R_2}
 \end{array}
 \right),
\end{align}
such that ${e^A}_{\mu}{e^{\star}_A}^{\nu}={\delta_{\mu}}^{\nu}$ and ${e^A}_{\mu}{e^{\star}_B}^{\mu}={\delta^A}_B$.
Here $R_0$, $R_1$, and $R_2$ are the radii for the directions $\tau$, $x$, and $y$, and $\alpha$, $\beta$, and $\gamma$
are related to the angles between directions $\tau$ and $x$,  $x$ and $y$, and $\tau$ and $y$, respectively.
The Euclidean metric is then given by

\begin{align}
 g_{\mu\nu}
 &=
 {e^A}_{\mu}{e^B}_{\nu}\delta_{AB}
 \nonumber\\
 &=
  \left(
\begin{array}{ccc}
 R_0^2+\alpha^2R_1^2+\gamma^2R_2^2 & -\alpha R_1^2+\beta\gamma R_2^2 & -\gamma R_2^2 \\
 -\alpha R_1^2+\beta\gamma R_2^2 & R_1^2+\beta^2R_2^2 & -\beta R_2^2 \\
 -\gamma R_2^2 & -\beta R_2^2 & R_2^2
 \end{array}
 \right),
\end{align}
and the line element is
\begin{align}
 ds^2
 &=
 g_{\mu\nu}d\theta^{\mu}d\theta^{\nu}
 \nonumber\\
 &=
 R_0^2(d\theta^0)^2+R_1^2(d\theta^1-\alpha d\theta^0)^2
 + R_2^2(d\theta^2-\beta d\theta^1-\gamma d\theta^0)^2,
\end{align}
where $0\leq \theta^{\mu}\leq 2\pi$ are angular variables.

The group ${SL}(3,\mathbb{Z})$ is generated by two transformations:
\begin{align}
 U_1 =
 \left(
 \begin{array}{ccc}
 0 & 0 & 1 \\
 1 & 0 & 0 \\
 0 & 1 & 0
 \end{array}
\right),
\quad
 U_2 =
 \left(
 \begin{array}{ccc}
 1 & 1 & 0 \\
 0 & 1 & 0 \\
 0 & 0 & 1
 \end{array}
\right).
\label{generator}
\end{align}
An ${SL}(3,\mathbb{Z})$ transformation acts as follows on
the dreibein and metric:
\begin{align}
 {e^A}_{\mu}
 &\overset{L}{\longrightarrow}
 \ {{( e L^T)}^A}_{\mu}
 = {L_{\mu}}^{\rho} {e^A}_{\rho},
 \nonumber\\
 {e^{\star}_A}^{\mu}
 &\overset{L}{\longrightarrow}
 {{(e^{\star}L^{-1})}_A}^{\mu}
 =  {e^{\star}_A}^{\rho}{{(L^{-1})}_{\rho}}^{\mu},
 \nonumber\\
 g_{\mu\nu}
 &\overset{L}{\longrightarrow}
 \ {( L g L^T)}_{\mu\nu}
 = {L_{\mu}}^{\rho} {L_{\nu}}^{\sigma}  g_{\rho\sigma},
\end{align}
for any ${SL}(3,\mathbb{Z})$ element $ L = U^{n_1}_1 U^{n_2}_2 U^{n_3}_1\cdots$.
Under the $U_2$ transformation,
the metric transforms according to
\begin{align}
& g_{\mu\nu}
 \overset{U_2}{\longrightarrow}
 \  {( U_2 g U^T_2)}_{\mu\nu}
 =
 \left(
  \begin{matrix}
 R_0^2+(\alpha-1)^2R_1^2+(\gamma+\beta)^2R_2^2 & -(\alpha-1) R_1^2+\beta(\gamma+\beta) R_2^2 & -(\gamma+\beta) R_2^2 \\
 -(\alpha-1) R_1^2+\beta(\gamma+\beta) R_2^2 & R_1^2+\beta^2R_2^2 & -\beta R_2^2 \\
 -(\gamma+\beta) R_2^2 & -\beta R_2^2 & R_2^2
 \end{matrix}
 \right),
\end{align}
which corresponds to the changes
\begin{align}
\alpha \rightarrow \alpha-1, \quad
\gamma \rightarrow \gamma+\beta,
\end{align}
while $R_0$, $R_1$, $R_2$, and $\beta$ are unchanged.

On the other hand, the less trivial generator $U_1$ can be decomposed as
\begin{align}
\label{U1' and M}
 U_1
 &=
 U'_1M,
 \quad
 U'_1
=
 \left(
 \begin{array}{ccc}
 0 & -1 & 0 \\
 1 & 0 & 0 \\
 0 & 0 & 1
 \end{array}
\right)
\quad
 M =
 \left(
 \begin{array}{ccc}
 1 & 0 & 0 \\
 0 & 0 & -1 \\
 0 & 1 & 0
 \end{array}
\right)
\end{align}
where $U_1^{\prime}$ corresponds to the $90^{\circ}$ rotation in the $\tau-x$ plane and $M$ is the $90^{\circ}$ rotation in the $x-y$ plane. The generator $U^{\prime}_1$ acts on the metric as
\begin{align}
& g_{\mu\nu}
 \overset{U'_1}{\longrightarrow}
 \  {( U'_1 g U'^T_1)}_{\mu\nu}
 =
 \left(
  \begin{array}{ccc}
 R_1^2+\beta^2R_2^2 & \alpha R_1^2-\beta\gamma R_2^2 & \beta R_2^2 \\
 \alpha R_1^2-\beta\gamma R_2^2 & R_0^2+\alpha^2R_1^2+\gamma^2R_2^2 & -\gamma R_2^2 \\
 \beta R_2^2 & -\gamma R_2^2 & R_2^2
 \end{array}
 \right),
\end{align}
which corresponds to the changes
\begin{align}
&R_0 \rightarrow R_0/|{\tau}|, \quad
R_1 \rightarrow R_1|{\tau}|, \quad
\tau_1 \rightarrow -\tau_1/|{\tau}|^2,
\quad
\gamma \rightarrow -\beta, \quad
\beta \rightarrow \gamma \quad
\text{(while $R_2$ is unchanged)},
\end{align}
where we have introduced
\begin{align}
{\tau}\equiv\alpha+ir_{01},
\quad
r_{01}\equiv R_0/R_1.
\end{align}
Observe also that under $R_0\to R_0/|\tau|$ and $R_1\to R_1|\tau|$, $\tau_2\to \tau_2/|\tau|^2$.
Hence, $U'_1$ induces $\tau \to -1/\tau$.

Finally, the transformation $M$ acts on the metric as
\begin{align}
 g_{\mu\nu}
 &
 \overset{M}{\longrightarrow}
 \  {( M g M^T)}_{\mu\nu}
=
 \left(
  \begin{array}{ccc}
 R_0^2+\alpha^2R_1^2+\gamma^2R_2^2 & \gamma R_2^2 & -\alpha R_1^2+\beta\gamma R_2^2 \\
 \gamma R_2^2 & R_2^2 & \beta R_2^2 \\
 -\alpha R_1^2+\beta\gamma R_2^2 & \beta R_2^2 & R_1^2+\beta^2R_2^2
 \end{array}
 \right).
\end{align}
The two transformations $U'_1$ and $U_2$ correspond respectively to
modular $S$ and $T^{-1}$ transformations in the ${\tau}-x$ plane, generating the ${SL}(2,\mathbb{Z})$ subgroup of ${SL}(3,\mathbb{Z})$ group. Combined with $M$, they generate the whole ${SL}(3,\mathbb{Z})$ group.
In the following, we denote $U^{\prime}_1 M$ by $\mathcal{S}$
and $U_2$ by $\mathcal{T}^{-1}$.

\section{Transformation properties of the characters
in the gauged $\mathbb{Z}_{\mathrm{K}}\times \mathbb{Z}_{\mathrm{K}}$
gauge theory}
\label{transformation laws}

In this appendix, 
we list 
the transformation properties of 
the characters of the topological 
$\mathbb{Z}_{\mathrm{K}}\times \mathbb{Z}_{\mathrm{K}}$
gauge theory after gauging the $\mathbb{Z}_2$ symmetry.
From these 
transformation properties, one can construct the 
$\mathcal{S}$ matrix. 
\begin{align}
U_1^{\prime}\chi_{n_i}^0&=\frac{1}{2\mathrm{K}^2}\sum_{n_0^{\prime}n_1^{\prime}}e^{\frac{2\pi i}{\mathrm{K}}(2n_0n_1^{\prime}+2n_1n_0^{\prime})}(\chi_{n_0^{\prime},n_1^{\prime},n_2}^0+\chi_{n_0^{\prime},n_1^{\prime},n_2}^1)
+
\frac{1}{\mathrm{K}^2}\sum_{n_{0}^{\prime},n_1^{\prime},l_{0}^{\prime},l_1^{\prime}}
e^{\frac{2\pi i}{\mathrm{K}}[n_0(n_1^{\prime}+l^{\prime}_1)+n_1(n_0^{\prime}+l^{\prime}_0)]}\chi_{n_0^{\prime},n_1^{\prime},n_2}^{l_0^{\prime},l_1^{\prime},l_2}
\nonumber\\
&\quad
+\frac{1}{2\mathrm{K}}\sum_{m_0^{\prime},m_1^{\prime}}e^{\frac{2\pi i}{\mathrm{K}}(n_0m_1^{\prime}+n_1m_0^{\prime})}(\chi_{\sigma_x}^{m_0^{\prime},m_1^{\prime},n_2}+\chi_{\tau_x}^{m_0^{\prime},m_1^{\prime},n_2}),
\nonumber \\
U_1^{\prime}\chi_{n_i}^{l_i}&=\frac{1}{\mathrm{K}^2}\sum_{n_0^{\prime},n_1^{\prime}}
e^{\frac{2\pi i}{\mathrm{K}}[n_0^{\prime}(n_1+l_1)+n_1^{\prime}(n_0+l_0)]}(\chi_{n_{0,1}^{\prime},n_2}^0+\chi_{n_{0,1}^{\prime},n_2}^1)
+
\frac{1}{\mathrm{K}^2}\sum_{n_0^{\prime},n_1^{\prime},l_0^{\prime},l_1^{\prime}}e^{\frac{2\pi i}{\mathrm{K}}(n_0n_1^{\prime}+n_1n_0^{\prime}+l_0l_1^{\prime}+l_1l_0^{\prime})}\chi_{n_0^{\prime},n_1^{\prime},n_2}^{l_0^{\prime},l_1^{\prime},l_2},
\nonumber \\
U_1^{\prime}\chi_{\sigma_x}^{m_i}
&=\frac{1}{2\mathrm{K}}\sum_{m_0^{\prime},m_1^{\prime}}e^{\frac{2\pi i}{\mathrm{K}}(m_0m_1^{\prime}+m_1m_0^{\prime})}(\chi_{m_{0,1}^{\prime},m_2}^0-\chi_{m_{0,1}^{\prime},m_2}^1)
\nonumber\\
&\quad
+\frac{1}{4\mathrm{K}}\sum_{m_0^{\prime},m_1^{\prime}}\left[ 1+(-1)^{m_1+m_1^{\prime}}+(-1)^{m_0+m_0^{\prime}}(1+(-1)^{m_1+m_1^{\prime}+\mathrm{K}}) \right]
e^{\frac{\pi i}{\mathrm{K}}(m_0m_1^{\prime}+m_1m_0^{\prime})}(\chi_{\sigma_x}^{m_0^{\prime},m_1^{\prime},m_2}-\chi_{\tau_x}^{m_0^{\prime},m_1^{\prime},m_2}),
\nonumber \\
U_1^{\prime}\chi_{\sigma_y}^{m_i}&=\frac{1}{2\mathrm{K}}\sum_{m_0^{\prime},m_1^{\prime}}e^{\frac{2\pi i}{\mathrm{K}}(m_0m_1^{\prime}+m_1m_0^{\prime})}\left( \chi_{\sigma_y}^{m_0^{\prime},m_1^{\prime},m_2}+\chi_{\tau_y}^{m_0^{\prime},m_1^{\prime},m_2} \right)
+
\frac{1}{2\mathrm{K}}\sum_{m_0^{\prime},m_1^{\prime}}e^{\frac{2\pi i}{\mathrm{K}}(m_0m_1^{\prime}+m_1m_0^{\prime})}\left( \chi_{\sigma_y}^{m_0^{\prime},m_1^{\prime},m_2}+\chi_{\tau_y}^{m_0^{\prime},m_1^{\prime},m_2} \right),
\nonumber \\
U_1^{\prime}\chi_{\sigma_{xy}}^{m_i}&=\frac{1}{2\mathrm{K}}\sum_{m_0^{\prime},m_1^{\prime}}
e^{\frac{2\pi i}{\mathrm{K}}(m_0m_1^{\prime}+m_1m_0^{\prime})}\left( \chi_{\sigma_{xy}}^{m_0^{\prime},m_1^{\prime},m_2}+\chi_{\tau_{xy}}^{m_0^{\prime},m_1^{\prime},m_2} \right)
+\frac{1}{2\mathrm{K}}\sum_{m_0^{\prime},m_1^{\prime}}e^{\frac{2\pi i}{\mathrm{K}}(m_0m_1^{\prime}+m_1m_0^{\prime})}\left( \chi_{\sigma_{xy}}^{m_0^{\prime},m_1^{\prime},m_2}-\chi_{\tau_{xy}}^{m_0^{\prime},m_1^{\prime},m_2} \right). 
\end{align}
\end{widetext}

\bibliography{reference}

\end{document}